\documentstyle[epsfig]{mn}
\newif\ifAMStwofonts

\ifoldfss
  \ifCUPmtlplainloaded \else
    \NewTextAlphabet{textbfit} {cmbxti10} {}
    \NewTextAlphabet{textbfss} {cmssbx10} {}
    \NewMathAlphabet{mathbfit} {cmbxti10} {} 
    \NewMathAlphabet{mathbfss} {cmssbx10} {} 
  \fi
  \ifAMStwofonts
    \ifCUPmtlplainloaded \else
      \NewSymbolFont{upmath} {eurm10}
      \NewSymbolFont{AMSa} {msam10}
      \NewMathSymbol{\upi}     {0}{upmath}{19}
      \NewMathSymbol{\umu}     {0}{upmath}{16}
      \NewMathSymbol{\upartial}{0}{upmath}{40}
      \NewMathSymbol{\leqslant}{3}{AMSa}{36}
      \NewMathSymbol{\geqslant}{3}{AMSa}{3E}

      \let\leq=\leqslant \let\le=\leqslant
      \let\geq=\geqslant 
    \fi
  \fi
\fi 

\ifnfssone
  \newmathalphabet{\mathit}
  \addtoversion{normal}{\mathit}{cmr}{m}{it}
  \addtoversion{bold}{\mathit}{cmr}{bx}{it}
  \newmathalphabet{\mathbfit} 
  \addtoversion{normal}{\mathbfit}{cmr}{bx}{it}
  \addtoversion{bold}{\mathbfit}{cmr}{bx}{it}
  \newmathalphabet{\mathbfss} 
  \addtoversion{normal}{\mathbfss}{cmss}{bx}{n}
  \addtoversion{bold}{\mathbfss}{cmss}{bx}{n}
  \ifAMStwofonts
    \ifCUPmtlplainloaded \else
      \UseAMStwoboldmath
      \makeatletter
      \new@mathgroup\upmath@group
      \define@mathgroup\mv@normal\upmath@group{eur}{m}{n}
      \define@mathgroup\mv@bold\upmath@group{eur}{b}{n}
      \edef\UPM{\hexnumber\upmath@group}
      \new@mathgroup\amsa@group
      \define@mathgroup\mv@normal\amsa@group{msa}{m}{n}
      \define@mathgroup\mv@bold\amsa@group{msa}{m}{n}
      \edef\AMSa{\hexnumber\amsa@group}
      \makeatother
      \mathchardef\upi="0\UPM19
      \mathchardef\umu="0\UPM16
      \mathchardef\upartial="0\UPM40
      \mathchardef\leqslant="3\AMSa36
      \mathchardef\geqslant="3\AMSa3E

      \let\leq=\leqslant \let\le=\leqslant
      \let\geq=\geqslant 
    \fi
  \fi
\fi 

\ifnfsstwo
  \DeclareMathAlphabet{\mathbfit}{OT1}{cmr}{bx}{it}
  \SetMathAlphabet\mathbfit{bold}{OT1}{cmr}{bx}{it}
  \DeclareMathAlphabet{\mathbfss}{OT1}{cmss}{bx}{n}
  \SetMathAlphabet\mathbfss{bold}{OT1}{cmss}{bx}{n}
  \ifAMStwofonts
    \ifCUPmtlplainloaded \else
      \DeclareSymbolFont{UPM}{U}{eur}{m}{n}
      \SetSymbolFont{UPM}{bold}{U}{eur}{b}{n}
      \DeclareSymbolFont{AMSa}{U}{msa}{m}{n}
      \DeclareMathSymbol{\upi}{0}{UPM}{"19}
      \DeclareMathSymbol{\umu}{0}{UPM}{"16}
      \DeclareMathSymbol{\upartial}{0}{UPM}{"40}
      \DeclareMathSymbol{\leqslant}{3}{AMSa}{"36}
      \DeclareMathSymbol{\geqslant}{3}{AMSa}{"3E}

      \let\leq=\leqslant \let\le=\leqslant
      \let\geq=\geqslant 
    \fi
  \fi
\fi 

\ifCUPmtlplainloaded \else
  \ifAMStwofonts \else 
    \def\upi{\pi}
    \def\umu{\mu}
    \def\upartial{\partial}
  \fi
\fi

\title[Distant field BHB stars and the mass of the Galaxy II] {Distant
field BHB stars and the mass of the Galaxy II: Photometry and
spectroscopy of UKST candidates $16<B<19.5, 11<R<52$ kpc \thanks{Based
on observations obtained at the Jacobus Kapteyn Telescope, the Isaac
Newton Telescope, and the William Herschel Telescope, La Palma, 
the Anglo-Australian Telescope, and the ANU 2.3-m telescope, Siding Spring
Observatory, Australia.}}

\author[L.Clewley, et al.]
{L. Clewley,$^{1}$ S. J. Warren,$^{2}$ P. C. Hewett,$^{3}$
John E. Norris$^{4}$, N.W. Evans$^{3}$\\
$^{1}$Department of Physics, Denys Wilkinson Bldg., University of
Oxford, Keble Road, Oxford, OX1 3RH \\
$^{2}$Blackett Laboratory, Imperial College of Science Technology
and Medicine, Prince Consort Rd, London SW7 2BW \\
$^{3}$Institute of Astronomy, Madingley Road, Cambridge CB3 0HA\\
$^{4}$Research School of Astronomy \& Astrophysics, The Australian National
University, Mount Stromlo Observatory,\\ Cotter Road, Weston, ACT 2611,
Australia\\}
\date{Accepted
      Received
      in original form}

\begin{document}
\maketitle

\begin{abstract}
This is the second in a series of papers presenting a new calculation
of the mass of the Galaxy based on radial velocities and distances for
a sample of faint $16<B<21.3$ field blue horizontal-branch (BHB)
stars. We present accurate $BV$ CCD photometry and spectra for 142
candidate A--type stars selected from $ub_jr$ photometry of UK Schmidt
telescope plates in six high--Galactic--latitude fields.
Classification of these candidates produces a sample of 60 BHB stars
at distances of 11--52 kpc from the Sun (mean 28 kpc), with
heliocentric line-of-sight velocities accurate to 15 km\,s$^{-1}$, and
distance errors $<10\%$. We provide a summary table listing
coordinates and velocities of these stars. The measured dispersion of
the radial component of the Galactocentric velocity for this sample is
108 $\pm$ 10 km s$^{-1}$, in agreement with a recent study of the
distant halo by Sirko and coworkers. Measurements of the Ca II K line
indicate that nearly all the stars are metal-poor with a mean [Fe/H] =
-1.8 with dispersion 0.5.  Subsequent papers will describe a second
survey of BHBs to heliocentric distances $70<R<125$ kpc and present a
new estimate of the mass of the Galaxy.
\end{abstract}

\begin{keywords}
Galaxy:   halo  --   stars:   blue  horizontal-branch 
\end{keywords}

\section{Introduction} In contrast with the detailed knowledge we have of the
stellar content of the halo of the Galaxy, the dark matter content is
far less well characterised, with both the total mass and the size of
the halo poorly determined quantities.  The accurate measurement of
the mass profile would provide important clues to the nature of the
dark matter.  For instance, an accurate estimate of the mass within a
radius of 50 kpc is required to establish the fraction of the mass in
compact objects identified from microlensing experiments (e.g Alcock
et al. 2000).  Quantifying the distribution of dark matter in the
Galaxy is essential for understanding the assembly of the various
baryonic components through comparison with simulations.  This paper
is the second in a series of four, presenting a new measurement of the
mass profile of the Galaxy out to the largest radii,
$r>100$kpc,\footnote{In this paper we use the coordinate $r$ to denote
Galactocentric distances and the coordinate $R$ to denote heliocentric
distances} using surveys for remote halo blue horizontal branch (BHB)
stars.

The main shortcoming of previous analyses of the mass of the Galaxy is
the small size of the sample of objects at large radii, used as
dynamical tracers.  Wilkinson and Evans (1999, hereafter WE99)
calculate the total mass of the Milky Way to be
$M_{tot}=1.9^{+3.6}_{-1.7} \times 10^{12} M_{\odot}$, using the full
set of 27 known satellite galaxies and globular clusters at
Galactocentric radii $r>$ 20 kpc (six of which possess measured proper
motions).  This sample must be nearly complete, so a new population of
distant halo objects must be found in order to increase the number of
dynamical tracers.  Field BHB stars are ideal for this purpose.  These
A--type stars are luminous, $M_V=0.9$ (\S5), which ensures that they can be
detected to large distances, and have a small spread in absolute
magnitude, so their distances may be determined accurately.

In a recent paper Sakamoto et al. (2003) added new kinematic data to
the dataset used by WE99, and obtained a considerably more precise
estimate of the total mass of the Galaxy of $M_{tot}=2.5^{+0.5}_{-1.0}
\times 10^{12} M_{\odot}$.  Primarily, the new kinematic data comprise
radial velocities of 412 BHB stars, of which 211 have proper motions, at
heliocentric distances $R<10$kpc, with a median distance of $\sim$
4.5kpc. Sakamoto et al. adopted the WE99 mass model, where the total
mass enclosed within Galactocentric radius $r$ is
$M(r)=M_{tot}/(1+a^2/r^2)^{1/2}$.  Here $a$ is the scale length, for
which the best-fit value of 225kpc was obtained.  For $r=20$kpc, i.e.
a radius containing all the BHB stars used in the analysis, this gives
$M(r)=0.09M_{tot}$.  This indicates that most of the improved
precision derives from kinematic data within a radius containing only
one tenth of the total mass, and therefore relies on extrapolation of
a model that is tightly constrained only at small radii. In a review
of mass estimates of the Milky Way halo, Zaritsky (1999) emphasises
the dangers of such extrapolation (see also Bellazzini, 2004, for a
useful discussion). This motivates a new survey for remote BHB stars, at
distances that are a substantial fraction of the halo scale length
$a$, in order to obtain a direct measure of the enclosed mass at large
radii.

Although BHB stars are abundant in the Galaxy halo, selection of a
clean sample is not straightforward.  Samples of field A--type stars
in the halo are easily identifiable in UBR (or equivalent e.g.  $ugr$)
multicolour datasets (e.g.  Yanny et al. 2000).  Samples selected by
broadband colour include not only BHB stars but also stars of main
sequence surface gravity that are some 2 magnitudes less luminous,
predominantly field blue stragglers (hereafter A/BS), as well as a
small proportion of quasars.  The reason that no large sample of
remote $r>30$kpc BHB stars has yet been compiled, is that the methods
developed for separating the two populations of A stars (e.g.  Kinman,
Suntzeff, and Kraft, 1994) require signal-to-noise ($S/N$) ratios that
are unfeasibly high for such faint stars ($B>18$).  In the first paper
in this series (Clewley et al. 2002, hereafter Paper I), we developed
two methods that overcome the difficulties.  The methods require
relatively modest telescope resources, yet produce samples with high
completeness and low contamination.  The methods are applicable
specifically to the classification of halo stars with strong Balmer
lines, defined by EW $H\gamma>13$\AA, i.e.  A stars in the approximate
colour range $0.0<(B-V)_0<0.2$.

Both methods employ a Sersic function fit to the H$\gamma$ and
H$\delta$ absorption lines.  The first method, the {\em
$D_{0.15}$--Colour} method, plots the average of the width of the two
Balmer lines against $(B-V)_0$ colour.  The A/BS stars, having higher
surface gravity, separate from the BHB stars because of their broader
Balmer lines.  The EW of the CaK line is used to filter out a small
number of interlopers in the BHB sample.  We used Monte Carlo methods
to establish that with $(B-V)_0$ colours accurate to $0.03$
magnitudes, and spectra of $S/N=15{\rm\AA}^{-1}$, samples of BHB stars
selected by this method would be about $87\%$ complete, with a
contamination of $7\%$ by A/BS stars.  Contamination at this level can
safely be accounted for in the dynamical analysis.  Spectra of this
$S/N$ are also suitable for measurement of the radial velocities.  The
second method, called the {\em Scale width--Shape} method, plots two
parameters of the Sersic fit, the scale width $b$, against the
power--law exponent $c$.  The method is almost as efficient as the
{\em $D_{0.15}$--Colour} method, with $82\%$ completeness and $12\%$
contamination, for the same spectroscopic $S/N$ of $15{\rm\AA}^{-1}$.
Again, the EW of the CaK line is used to filter out a small number of
interlopers.  The advantage of the {\em Scale width--Shape} method is
that colours are not needed.  For samples of stars with existing
accurate $(B-V)_0$ colours, the first method is preferred, as the
contamination is lower, and the completeness higher.  The gain is
small, however, and accurate photometry is time consuming.  Where
accurate colours are not already available, if telescope resources are
limited, the best practical solution will be simply to obtain spectra,
and use the {\em Scale width--Shape} method.

In this second paper we describe the compilation of a sample of BHB
stars in the magnitude range $16<B<19.5$, corresponding to heliocentric
distances $11<R<52$ kpc, using the two classification methods
described above.  This first survey (hereafter the UKST survey) starts
from photographic plates taken with the United Kingdom Schmidt
Telescope (UKST).  In the third paper we describe a second survey, for
fainter BHB stars, $20<g*<21.1$, corresponding to heliocentric
distances $70<R<125$ kpc.  This second survey starts from the Sloan
Digital Sky Survey (SDSS) Early Data Release (EDR) (Stoughton et
al. 2002).  The fourth paper will present the dynamical analysis of
these two data sets, combined with the 27 satellites and globular
clusters analysed by WE99.

The layout of the remainder of the paper is as follows.  In section 2
we describe the selection of the UKST fields, the reduction of the
plate scans, and the production of the multicolour photometric
datasets.  We provide catalogues of a total of 461 objects with the
colours of A--type stars, which are candidate halo BHB stars.  Section
3 covers CCD $B-V$ photometry of 280 candidate BHB stars, and section
4 describes the spectroscopic observations of 156 candidates.  We
describe the measurement of the metallicity and the radial velocity
from the spectra.  Section 5 reviews the measurement of distances of
BHB stars.  In Section 6 we use these results and the methods
developed in Paper I to classify these stars, and we provide a table
of distances and radial velocities of the 60 confirmed BHB stars.
Finally, in Section 7 we provide a summary of the main conclusions of
the paper.

\section{The selection of BHB candidates}

In this section we explain how the six UKST fields for the survey were
chosen, and we describe the compilation of calibrated three-band
photographic photometric datasets in each field.  We detail the
(dereddened) magnitude and colour criteria for the selection of
candidate BHB stars from these datasets, and provide complete lists of
the candidates for each field.  These candidate lists are suitable,
for example, for quantifying number counts of A--type stars in
different directions through the Galaxy.  On the other hand, the
sub-set of objects for which CCD photometry and spectroscopy were
obtained was not selected randomly from the candidate lists.  Rather,
both magnitude and colour criteria were applied.  For example, for the
CCD photometry, if the seeing was poor, brighter objects were selected
for observation, while spectroscopy was limited almost exclusively to
objects with confirmed CCD colours in the range $-0.05<(B-V)_0<0.25$,
since the classification methods work well only within this range. In
fact, because the colour selection criteria were not finalised at the
outset of the survey, a small number of objects that fall outside the
selection criteria used to define the complete candidate samples, were
also included in the CCD photometry and spectroscopy campaign. The
consequence of this is that the final list of confirmed BHB stars is
not a random sub-set of the complete candidate lists, and so cannot be
used, for example, for quantifying the density profile of BHB stars in
the halo.  On the other hand, since the stars have been selected
without reference to kinematic information, each confirmed BHB star
randomly samples the radial--velocity distribution function at that
distance, and so the sample is suitable for dynamical analysis.

\subsection{Field selection}

We selected the fields to give good coverage in opposing directions
above and below the Galactic plane, with a wide range of Galactic
longitudes.  Although the kinematic information is limited to radial
velocities, because of the offset of the Sun from the Galactic centre
the widely different lines of sight through the Galaxy provide some
complementary information on the anisotropy of the stellar orbits, and
also avoid the risk of the whole sample residing in a halo stream,
such as those associated with the Sagittarius dwarf galaxy (e.g.
Ibata, Gilmore \& Irwin, 1994, Dohm-Palmer et al. 2001, Newberg et
al. 2002).  The selection of fields was subject to the availability of
suitable plate material in the UKST archive.  This restricts the
search to negative declinations, and to fields containing pairs of
high--grade plates in the $u$, $b_j$, and $or/r$ bands.  We chose six
fields at high Galactic latitude, $|b| \geq 45^{\circ}$, to minimise
Galactic extinction. The coordinates of the centres of the six fields
are provided in Table \ref{tab_survey}, and plotted in Fig.  1.

\begin{figure}
\rotatebox{+90}{ \centering{ \scalebox{0.42}{
\includegraphics*[100,100][558,638]{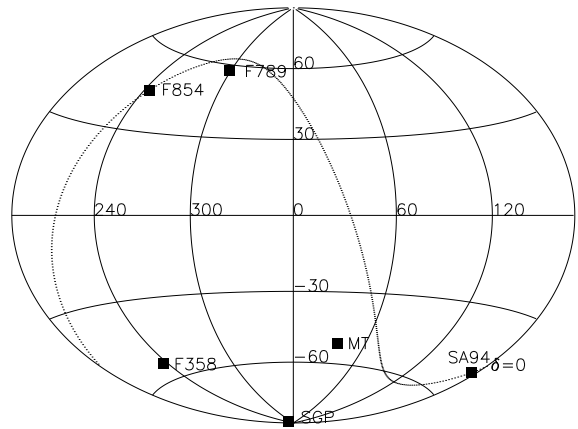} }}}
\caption{Galactic coordinate Aitoff projection showing the locations
of the six survey fields. The fields were chosen to sample different
directions through the Galaxy, within the constraints imposed by the
availability of suitable plate material in the UKST archive.
High Galactic-latitude fields were chosen to minimise extinction. The
dotted line plots zero declination.}
\label{plot_survey}
\end{figure}
\begin{table}
 \centering
  \begin{tabular}{@{}lrrrrrr}
   \hline \\[-12pt]
	Field name & \multicolumn{1}{c}{l} & \multicolumn{1}{c}{b} &
 \multicolumn{1}{c}{RA} & \multicolumn{1}{c}{Dec}\\
	           & & &
 \multicolumn{2}{c}{J 2000} \\
	\hline \\[-12pt]
SGP  & 250 &  -89  &    0 55  &  -27 47\\ 
SA94 & 175 &  -50  &    2 53  &    0 12\\ 
F358 & 236 &  -54  &    3 38  &  -34 50\\ 
F854 & 244 &   45  &   10 23  &   -0 15\\ 
F789 & 299 &   58  &   12 43  &   -5 16\\ 
MT   &  37 &  -51  &   22 06  &  -18 39\\ 
   \hline
\end{tabular}
\caption{Galactic and equatorial coordinates of the centres of six fields
observed in the survey, ordered by Right Ascension.\label{tab_survey}} 
\end{table}

\begin{table*}   
\begin{flushleft}
\begin{center}
\begin{tabular}{lccrrr}
\hline
\noalign{\smallskip}   
\multicolumn{1}{c}{Field} &   
\multicolumn{1}{c}{Area} &     
\multicolumn{1}{c}{Reddening} &   
\multicolumn{1}{c}{Sample $b_j$} &
\multicolumn{1}{c}{Sample $b_j$} & 
\multicolumn{1}{c}{No.} \\
\multicolumn{1}{c}{} &   
\multicolumn{1}{c}{(sq. deg.)} &      
\multicolumn{1}{c}{$\langle E(B-V) \rangle$} &      
\multicolumn{1}{c}{range (red)} &    
\multicolumn{1}{c}{range (unred)} &
\multicolumn{1}{c}{cand.}\\
\noalign{\smallskip}   
\hline  \\[-12pt]
SGP   &    24.8  &    0.018  &   $18.47-19.77$ &   $18.40-19.70$ &  64\\
MTF   &    25.6  &    0.032  &   $17.63-19.63$ &   $17.50-19.50$ & 128\\
SA94  &    25.7  &    0.067  &   $16.27-19.97$ &   $16.00-19.70$ &  72\\
F358  &    23.6  &    0.011  &   $17.34-19.74$ &   $17.30-19.70$ &  28\\
F789  &    25.3  &    0.029  &   $15.32-19.02$ &   $15.20-18.90$ & 114\\
F854  &    25.3  &    0.048  &   $16.19-19.59$ &   $16.00-19.40$ &  55\\   
\hline 
\end{tabular}
\vspace{2mm}
\caption{Data for the six fields in the survey. Listed are the
effective area of each field, the mean reddening, $\langle E(B-V)
\rangle$, the range of $b_j$ magnitudes with and without the reddening
correction and the total number of candidate stars in each
field.\label{tab_boxinfo}}
\end{center}
\end{flushleft}
\end{table*}

\subsection{Compilation of uncalibrated three-band photographic
photometric datasets} Pairs of photographic plates in each of the
three bands were scanned by the Automated Plate Measuring (APM)
machine (Kibblewhite et al. 1984) to produce lists of detected
objects, and their measured properties, including uncalibrated
fluxes. These lists were then matched and calibrated.

In four fields we used pairs of plates in the $u$, $b_j$, and $or$
pass-bands.  The exceptions were F358 which had a single $or$ plate,
and SA94 where we used two $r$ plates.  The $r$ band is a slightly
narrower and redder band than the standard $or$ band.  The effective
area of the scan in each field, approximately 25 square degrees, is
listed in Table \ref{tab_boxinfo}.  The effective area is the actual
area scanned, after allowing for regions drilled out around bright
stars, multiplied by a factor 0.8 to account for incompleteness due to
plate flaws, satellite trails and merging with neighbouring images.
The survey covers a total effective area of 150.3 square degrees over
the six fields.

In pairing and calibrating the object lists from the plates in each
field we followed the same reduction steps described in detail by
Warren at al. (1991) \---\ in fact the SGP catalogue described here is
the same catalogue used by them to search for quasars.  In outline,
the processing steps are as follows.  For each detected object the APM
measures a parameter {\em intensity} that scales monotonically and
non--linearly with flux, as well as several parameters describing the
light profile and the shape of each source.  The variation of the
light profile as a function of intensity, for stellar objects on the
plates, is used to calibrate the intensities onto a scale that is
approximately linear with flux (Bunclark \& Irwin, 1983).  The main
purpose of using pairs of plates in each band was to reduce the
photometric uncertainties by averaging the intensities.  The response
curve of the photographic emulsion is not perfectly uniform, but
varies across each plate (known as field effects).  In matching plates
taken in the same filter we gridded the area scanned, and calibrated
the intensity scale of the second plate to the scale of the first
plate in each patch i.e.  we forced the field effects on the second
plate to match those on the first plate.  A similar procedure was
adopted in order to match the object catalogues from different
passbands. The $b_j$ band was adopted as the reference passband and
any field effects in the other photometric bands were forced to match
any field effects present in the $b_j$ band.

We used the shape and light profile information to classify the
objects as stellar and non--stellar, combining the information from
all the plates in order to improve the discrimination.  Finally we
converted the intensities to a logarithmic scale and calibrated to
magnitudes using CCD observations of several stars on each plate, as
described below.

\subsection{Calibration}

For the calibration we made CCD observations in the $U$, $B$, and $R$
bands of selected fields.  We also used the CCD $BV$ photometry of the
candidates themselves (i.e.  all A-type stars).  We created $U$ and
$R$ sequences from the A--star $BV$ photometry using the relations:
$U-B = 0.5(B-V)$, for 0 $< (B-V) <$ 0.2; $B-R = 0.02 + 1.51(B-V)$ for
$(B-V) < 1.0$.  The $UBR$ photometry was then converted into the $u,
b_j, or, r$ natural system of the UKST using the relations,

\begin{eqnarray}
u = U - 0.01 - 0.01\,(U-B) \: \:\mathrm{for} \: -1.4 < U-B < 2.0,  \nonumber\\
b_{j} - B = 0.01 - 0.18\,(B-R) \:\: \mathrm{for} \: 0.0 < B-R < 1.6,   \nonumber\\
or - R = 0.00 - 0.01\,(B-R) \:\: \mathrm{for} \: 0.0 < B-R < 1.6,  \nonumber\\
r - R = -0.01 - 0.06\,(B-R) \:\: \mathrm{for} \: 0.0 < B-R < 1.6.  \nonumber
\end{eqnarray}

The number of stars which defined the calibration curve in any band
was typically 25.  The colour equations quoted above are either taken
from Warren et al. (1991), or they were computed using the procedures
described there.

Bearing in mind the earlier discussion of field effects, in
establishing the photometric errors there are two terms to consider.
There are random errors, which can be assessed for any band from the
scatter in a plot of magnitude difference between the two plates as a
function of magnitude (Warren et al. 1991, Fig.  3).  There are also
systematic errors, due to field effects on the reference plate in the
$b_j$ band, and these are the same in all passbands, due to the way in
which the field effects in all plates were forced to match those on
the reference $b_j$ plate.  These systematic errors do not contribute
to the uncertainties in the colours.  They are larger than the random
errors over the magnitude range of interest and were quantified by
measuring the scatter in the calibration curves.  Table
\ref{tab_schmidt_err} summarises the photometric uncertainties in the
photographic catalogues.

\begin{table}
\begin{flushleft}
\begin{center}
\begin{tabular}{cccc}
\hline
\noalign{\smallskip}   
 $b_j$ & $\sigma(b_j)$ & $\sigma(u-b_j)$ & $\sigma(b_j-or)$ \\
\noalign{\smallskip}   
\hline  \\[-12pt]
 16 & 0.11 & 0.05 & 0.05 \\
 17 & 0.11 & 0.05 & 0.05 \\
 18 & 0.11 & 0.05 & 0.05 \\
 19 & 0.11 & 0.06 & 0.06 \\
 20 & 0.11 & 0.09 & 0.10 \\
\hline 
\end{tabular}
\vspace{2mm}
\caption{Uncertainties for the photographic photometric data as a
function of $b_j$ magnitude. The reason why the magnitude errors are larger
than the colour errors is explained in the text.\label{tab_schmidt_err}}
\end{center}
\end{flushleft}
\end{table}
\begin{table*}   
\begin{flushleft}
\begin{center}
\begin{tabular}{rrrrcccc}
\hline
\noalign{\smallskip}   
\multicolumn{1}{c}{Star} &   
\multicolumn{1}{c}{R.A. (J2000.0)} &     
\multicolumn{1}{c}{Dec. (J2000.0)} &      
\multicolumn{1}{c}{$b_j$} &
\multicolumn{1}{c}{$u-b_j$} &
\multicolumn{1}{c}{$b_j-or$}&
\multicolumn{1}{c}{$\sigma(u-b_j)$} & 
\multicolumn{1}{c}{$\sigma(b_j-or)$} \\
\noalign{\smallskip}   
\hline\\[-12pt]   
CSGP-001  &  01 05 28.20 & -26 14 00.79 &  19.40 &   0.11 &  0.10  &  0.07 & 0.08\\
CSGP-002  &  01 03 35.32 & -25 54 49.64 &  19.05 &  -0.03 &  0.11  &  0.06 & 0.06\\
CSGP-003  &  01 02 44.21 & -26 27 46.75 &  18.87 &   0.16 &  0.34  &  0.06 & 0.06\\
CSGP-004  &  01 02 35.56 & -26 12 39.58 &  18.52 &   0.08 &  0.45  &  0.06 & 0.06\\
CSGP-005  &  01 01 56.71 & -25 38 03.13 &  18.63 &   0.16 &  0.19  &  0.06 & 0.06\\
CSGP-006  &  01 01 44.02 & -26 58 32.59 &  19.55 &   0.07 &  0.31  &  0.08 & 0.08\\
CSGP-007  &  00 59 23.47 & -25 48 36.96 &  19.56 &   0.20 &  0.22  &  0.08 & 0.08\\
\hline 
\end{tabular}
\vspace{2mm}
\caption{Photometric catalogue of A-type stars selected from six UKST
  fields shown in Figure \ref{two_col_plots}. The photometric
  uncertainties in these estimates are interpolated from those in Table
  \ref{tab_schmidt_err}. This table is presented in its entirety in
  the electronic edition of Monthly Notices. A portion is shown here
  for guidance regarding its content and form.\label{phot_schmidt}}
\end{center}
\end{flushleft}
\end{table*}

\begin{figure*}
\begin{minipage}{154mm}
\epsfig{figure=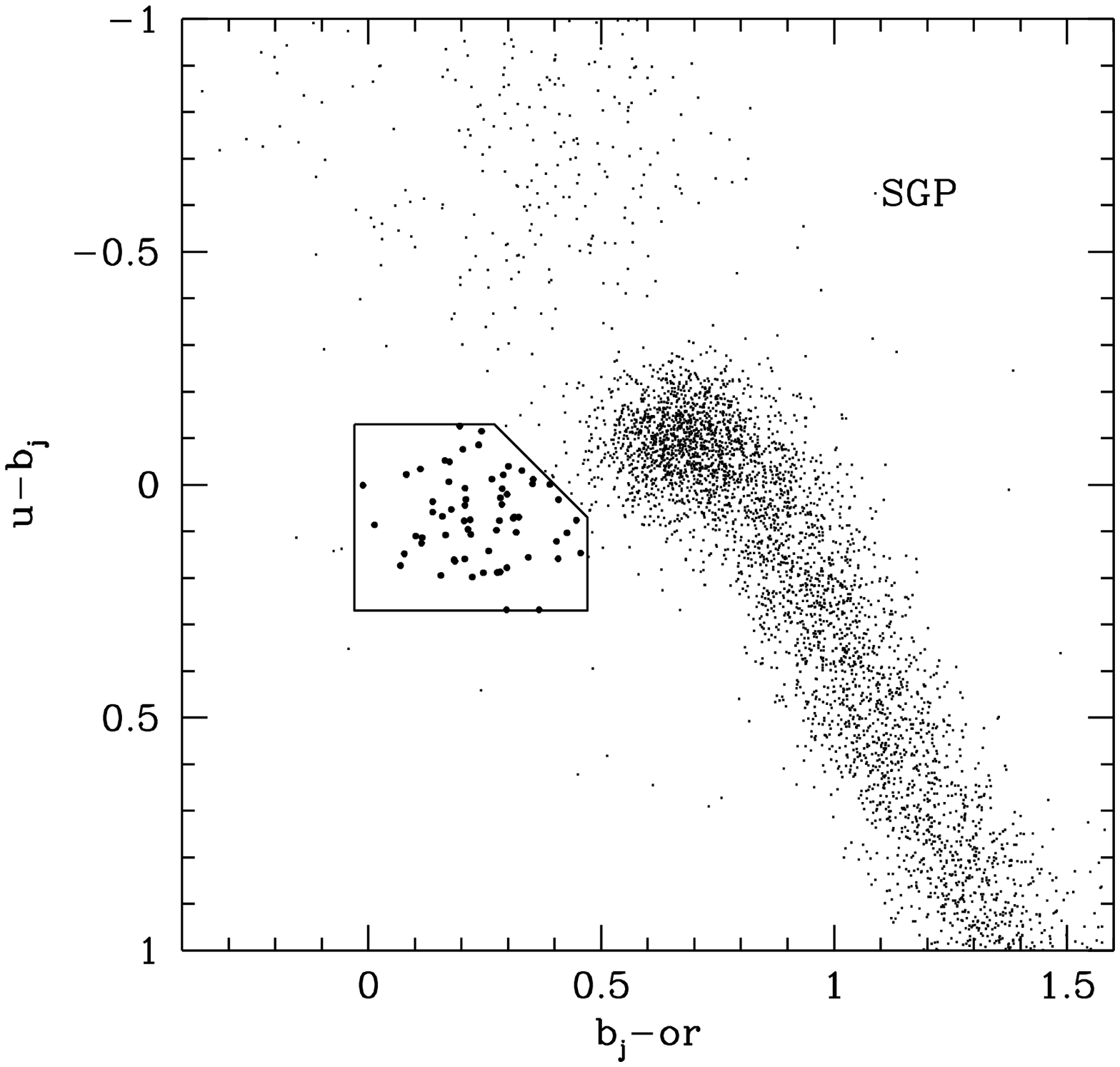,height=76mm,width=76mm}
\epsfig{figure=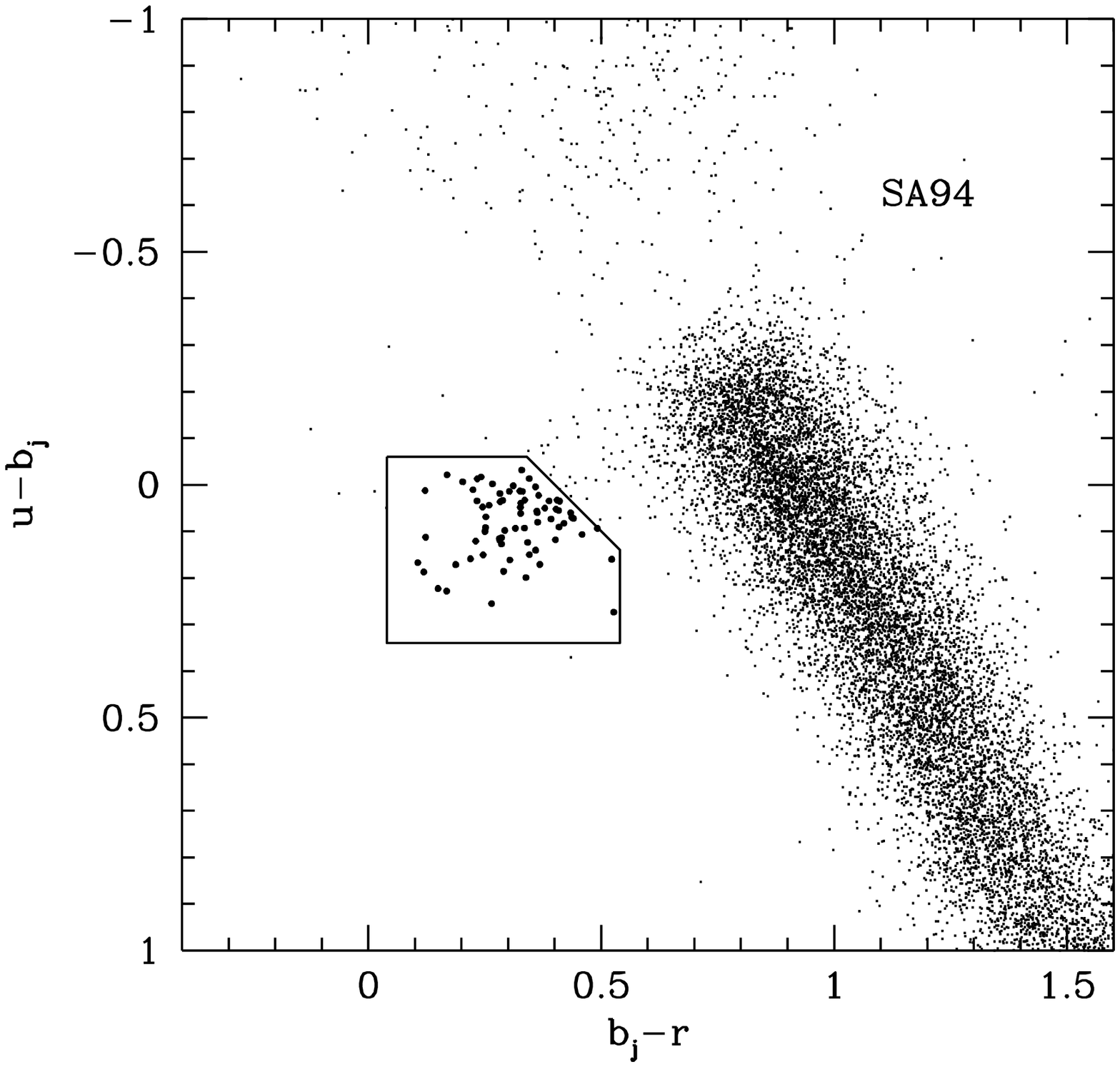,height=76mm,width=76mm}
\epsfig{figure=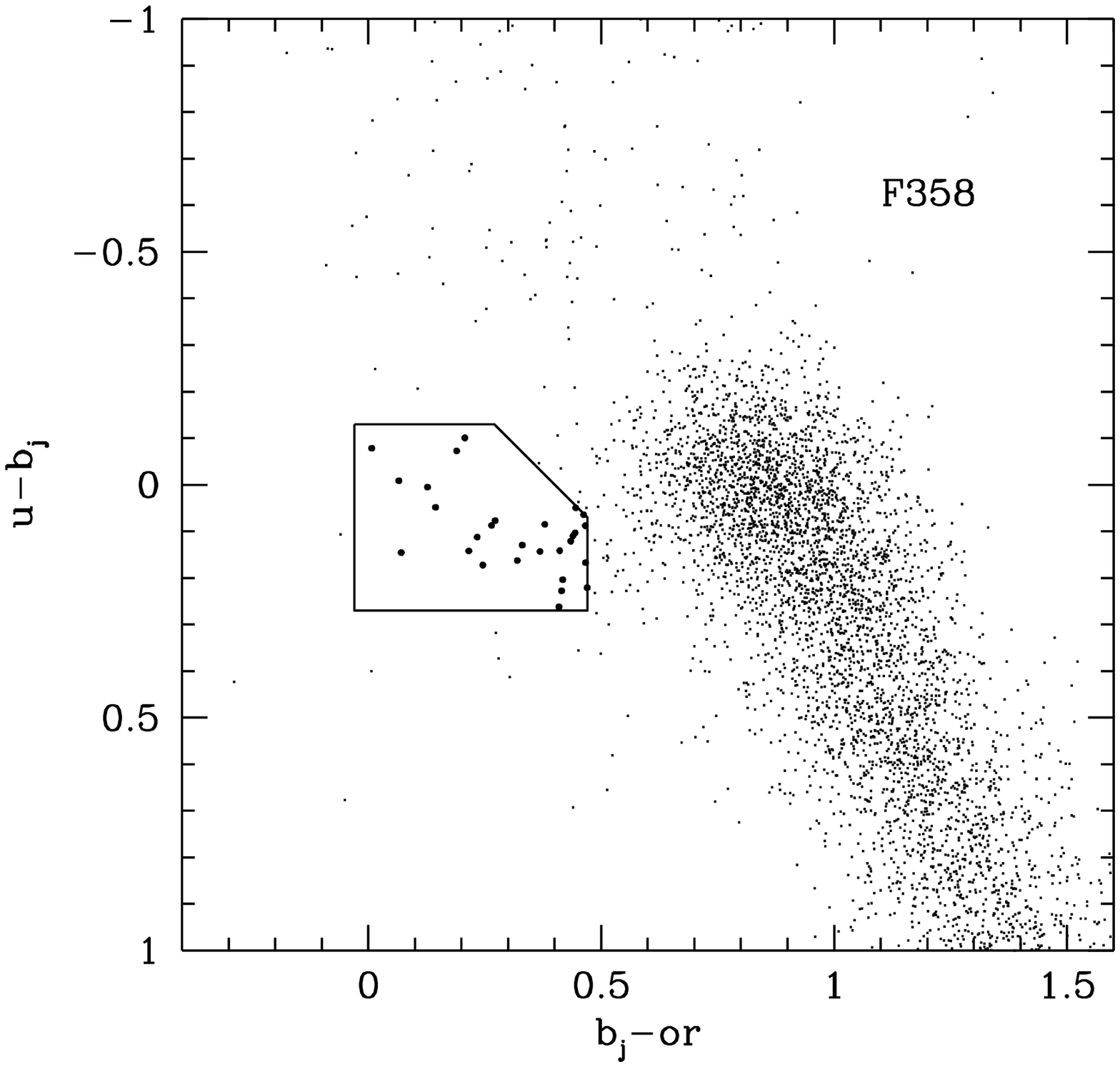,height=76mm,width=76mm}
\epsfig{figure=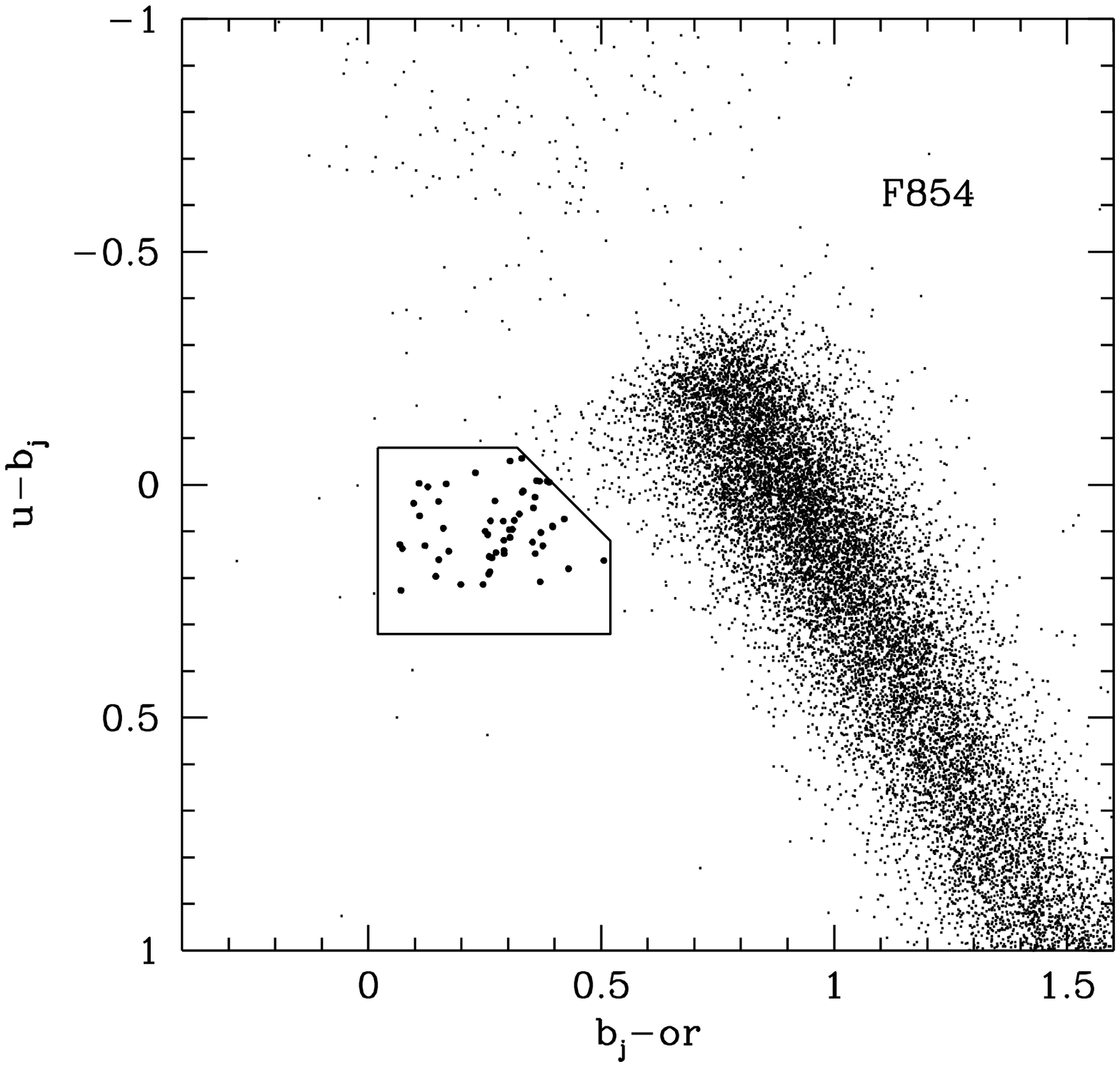,height=76mm,width=76mm}
\epsfig{figure=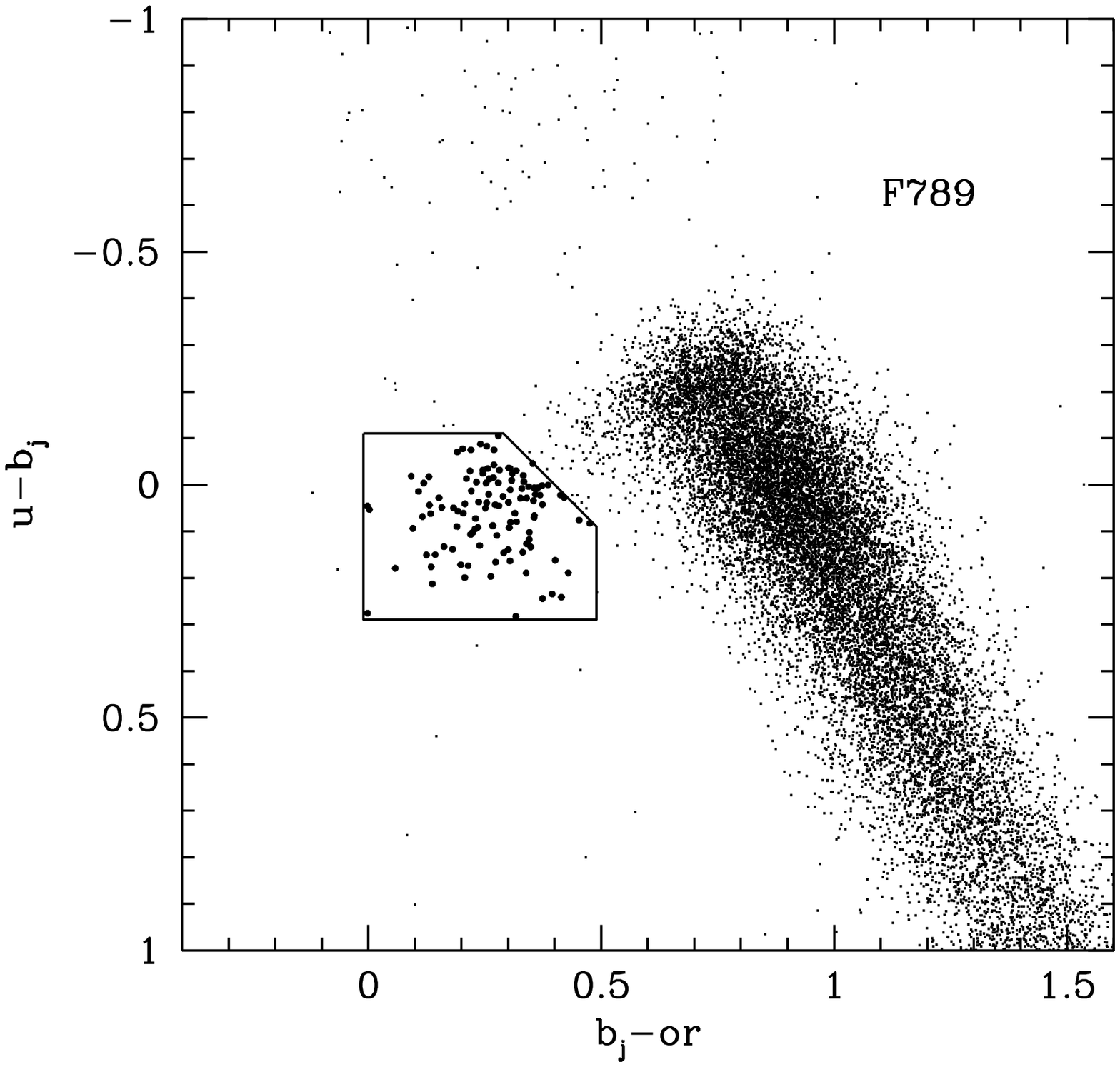,height=76mm,width=76mm}
\epsfig{figure=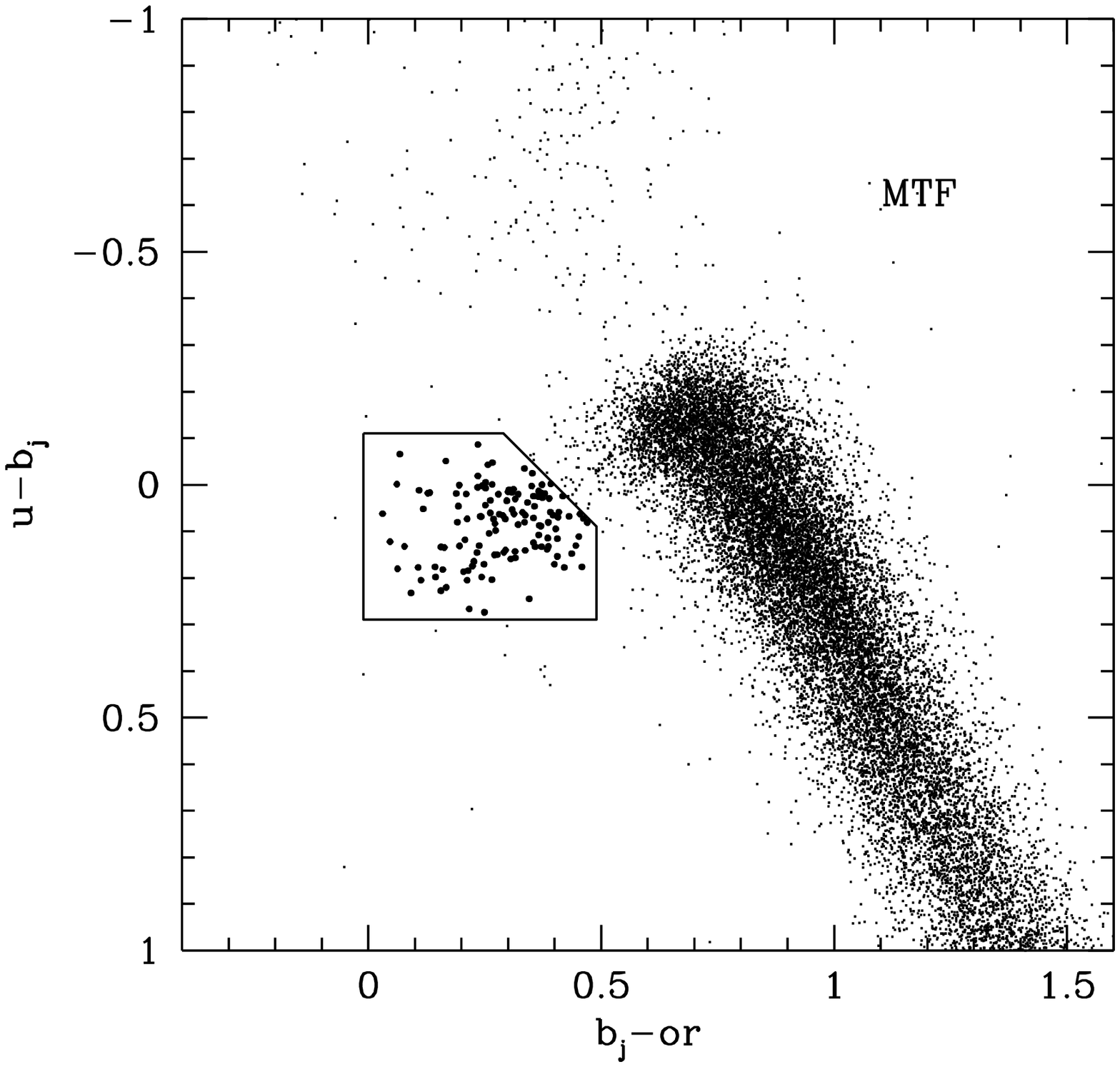,height=76mm,width=76mm}
\end{minipage}
\caption{Two-colour diagrams for the stellar targets for the six
fields listed in Table \ref{tab_survey}. The selection box defines the
complete sample of candidate BHB stars. The number of stars in each
box, their magnitude range and reddening correction are listed in Table
\ref{tab_boxinfo}. \label{two_col_plots}}
\end{figure*}

\subsection{Candidate selection}

Figure 2 shows two--colour $u-b_j$, $b_j-or/r$ diagrams for the
stellar objects in each field, and the selection boxes which define
the samples of candidate BHB stars.  This box is slightly different
for each field because we have allowed for Galactic reddening, since
the targets are at large distances. In considering candidate
selection, the $r$ band is treated as identical to the $or$ band. This
is because A stars have colours close to zero, so the difference
between the $r$ and $or$ magnitudes for A stars, i.e. the product of a
small colour and a small colour term, may be neglected. Similarly the
the difference in the extinction correction is negligible, because the
reddening in these fields is small (Table 2), and the wavelength
difference between the bands is also small. In this subsection all
equations involving the $or$ band apply also to the $r$ band.

The selection box is defined by
the unreddened colours as follows:

\begin{eqnarray}
\hspace{2.5cm} -0.15<(u-b_j)_0<0.25,\nonumber\\
-0.05<(b_j-or)_0<0.45,\nonumber \\
(u-b_j)_0-(b_j-or)_0>-0.40\nonumber.
\end{eqnarray}
The $(u-b_j)_0$ cut eliminates objects with large ultraviolet excesses
from the sample, such as quasars or white dwarfs.  The $(b_j-or)_0$
cut minimises contamination by F stars.  An additional diagonal cut is
made because the colour box lies closest to the stellar sequence in
this corner.  Converting $b_j-or$ to $B-V$ using
$B-V=-0.02+0.80(b_j-or)$ the corresponding range in $B-V$ is
$-0.06<B-V<(0.18-0.34)$, where the range in the red limit is a
consequence of the diagonal cut (seen in Figure 2).  Since we are
interested in BHB stars, which have $0.0<B-V<0.20$, the incompleteness
introduced by the diagonal cut itself will be negligible.

For each field we shifted the selection box along the reddening vector
by an amount appropriate for the value of $E(B-V)$.  The mean
reddening values for each field, $\langle E(B-V) \rangle$, are
provided in Table \ref{tab_boxinfo}, measured from the extinction maps
of Schlegel, Finkbeiner \& Davis (1998).  We used the reddening
relations $E(u-b_j)=1.40E(B-V)$, and $E(b_j-or)=1.36E(B-V)$ (note that
the reddening vector is nearly diagonal).  The relation between
extinction and reddening for the $b_j$ band is $A(b_j)=4.035E(B-V)$.
The magnitude range of the candidates is different for each field and
is listed in Table \ref{tab_boxinfo} together with the number of
candidates. The total number of candidates is 461. A catalogue of the
candidate A-type stars containing their $b_j$ magnitude, $b_j-or$ and
$u-b_j$ colours and their associated errors is provided in Table
\ref{phot_schmidt}. The stars in this sample are prefixed with a `C',
standing for `complete' sample.

The photometric calibration observations were, in fact, obtained as
the survey progressed, and the candidate BHB stars had to be selected
before the calibration stage.  This was possible because the clump of
A-type stars is clearly defined, and because the APM intensities are
nearly linear with flux.  The consequence of this is that some stars
observed photometrically or spectroscopically fall just outside the
selection boxes. 

In F854 we noted a significant faint over-density of fainter ($b_j$ =
20.2) BHB candidate stars in the direction $l = 244^{\circ}$, $b =
42^{\circ}$.  We later realised these mark the location of the dwarf
spheroidal galaxy Sextans, which resides at a heliocentric distance of
83 kpc.  In fact the magnitude limit of our complete sample in this
field is 19.59, so the complete sample is not contaminated by Sextans.
But it highlights the effectiveness of using Schmidt plates to
discover BHB stars out to at least 80 kpc.

\section{Photometry}

\subsection{Observations}

The {\em $D_{0.15}$--Colour} method of classification requires $B-V$
colours accurate to 0.03 magnitudes, so the photographic photometry of
Table \ref{phot_schmidt} is insufficient. To achieve a higher degree
of photometric accuracy, $BV$ CCD observations were made of candidate
BHB stars drawn from Table \ref{phot_schmidt}. We made a total of 421
observations of 280 stars over 47 photometric nights during the period
1998 to 2001. As a rule, the redder objects in the selection boxes
shown in Fig. \ref{two_col_plots} were given lower priority for
observation, since this region will contain fewer objects in the
desired colour range $0<(B-V)_0<0.2$. The observing dates are listed
in Table \ref{tab_phot_obs}.  Three telescopes were used for the
photometry.  (i) The 1.0-m Jacobus Kapteyn Telescope (JKT) at La
Palma.  We used a TEK1 1024x1024 CCD detector for the 1998 observing
run, which was replaced by a SITe2 2048x2048 CCD for subsequent runs.
Each chip has a scale of 0.33'' pixel$^{-1}$.  (ii) The 2.5-m Isaac
Newton Telescope (INT) at La Palma, using Chip 4 of the Wide Field
Camera, which is an EEV42-80 4096x2048 CCD with a scale of 0.33''
pixel$^{-1}$.  (iii) The ANU 2.3-m Telescope at Siding Spring
observatory (SSO), Australia, using a SITe 1024x1024 CCD with a
pixel scale of 0.59'' pixel$^{-1}$.  Generally the brighter stars were
observed at the JKT and the fainter ones at the SSO and INT. Repeat
observations were made between telescopes, in order to establish the
external errors.

Throughout each night we observed typically 30 standard stars from
Landolt (1992).  The standards were chosen to span a large range in
colour, and were observed over a wide range of airmass. Extinction and
colour terms were measured every night.  The target A-type stars were
observed as near to the meridian as possible, typically at airmasses
$\leq$ 1.3.  The integration times for each exposure were adjusted,
dependent on the brightness of the star, the brightness of the sky,
and the seeing, with the aim of achieving uniform $S/N$ for all
objects.

\subsection{Data reduction, analysis, and results}

Data reduction techniques employed standard routines available in {\tt
IRAF} (V2.11)\footnote{IRAF is distributed by the National Optical
Astronomy Observatories, which are operated by the Association of
Universities for Research in Astronomy, Inc.  under cooperative
agreement with the National Science Foundation.} for bias, flat-field
correction, and the removal of cosmic rays. The data taken on the ANU
2.3-m SSO telescope was reduced and analysed using FIGARO. Twilight
sky exposures were used to flat-field the data, and a correction was
then applied to account for the large-scale gradient in the sky. The
correction frames were created by combining and smoothing uncrowded
flat-fielded image frames, from which the stars were
$\sigma-$clipped. This procedure will, in fact, only be correct if any
scattered light in the instrument illuminates the CCD uniformly. With
hindsight, it would have been best to quantify the scattered light,
using a procedure such as that suggested by Manfroid, Selman, and
Jones (2001). Systematic errors are discussed below.

Stellar aperture photometry was performed using the APPHOT package in
{\tt IRAF}. For the standards we chose an aperture radius equal to 6.5
times the measured image full-width-half-maximum (FWHM).  The seeing
was typically 1.2'' -- 1.6'', giving an aperture radius of $\sim$
7.8'' -- 10.4''. The sky level was measured in an annulus between
radii of 10 and 20 FWHM.  We solved for the nightly $B$ and $V$
extinction and colour coefficients. The typical RMS scatter about the
best-fit relations in both $B$ and $V$ filters was 0.015
magnitudes. This is noticeably larger than the Poisson errors, and
the excess scatter could be due to scattered light.

\begin{table}
\centering
  \begin{tabular}{@{}lllll}
   \hline \\[-12pt]
	Telescope &Date&N&Fields\\
	\hline \\[-12pt]
        SSO & 1998\, Sep 13-16   	 &4& SGP \\
	JKT & 1998\, Oct 15, 17-20	 &5&  SA94, MTF \\
	SSO & 1999\, Mar 12-18           &7& F854, F789, SGP\\
	JKT & 1999\, April 4, 7-9 	 &4&  F854, F789 \\
	SSO & 1999\, Oct 11-17  	 &7& SGP \\
	JKT & 1999\, Oct 30-31, Sep 1, 3-5 &6&  SA94, MTF\\
	SSO & 1999\, Nov 17-20  	 &4& F358 \\
	SSO & 2000\, Feb 2-4    	 &4&  F854, F789\\
	JKT & 2000\, Mar 23, 26, 28   	 &3& F854, F789\\
	INT & 2001\, Mar 29-30, Apr 1  	 &3& SA94, MTF \\
\hline
\end{tabular}
\vspace{5mm}
\caption{The dates and total number of photometric nights `N' for the
  photometric survey in each of the six fields presented in Table
  \ref{tab_survey}. Observations were made using the 1.0-m Jacobus Kapteyn Telescope (JKT), the 2.5-m
Isaac Newton Telescope (INT) and the 2.3-m Telescope at Siding Spring
Observatories (SSO).\label{tab_phot_obs}}
\end{table}

For the survey stars we selected an aperture radius of 2.3 times the
average FWHM of stars in each frame. This small radius is close to
optimum in terms of $S/N$. To correct to the large aperture used for
the standards, we averaged the aperture correction measured for $\sim$
5 bright, isolated, unsaturated stars in each image frame. The
uncertainty on the aperture correction is not negligible, being
typically 0.01 magnitudes, and was added in quadrature to the Poisson
errors for the survey stars. The resulting $BV$ magnitudes and errors are
found by inverting the transformation solutions.  The total internal
photometric errors for the programme stars with $16 < V < 20$
respectively are in the range 0.015 - 0.030 magnitudes in $V$ and
0.020 - 0.040 in $B-V$. To obtain an estimate of the size of
any systematic errors, due, for example, to scattered light, we have
looked at the scatter in the photometry of objects observed on
different nights, and on different telescopes.

\begin{table}
\centering
  \begin{tabular}{@{}lcccc}
   \hline \\[-12pt]
Range in V magnitude & $N_{obs}$ & $N_{stars}$ & $\sigma_{V}$ & $\sigma(B-V)$ \\
\hline \\[-12pt]
 15 $<$ V $<$ 16 & 24  & 6 & 0.021 & 0.025 \\
 16 $<$ V $<$ 17 & 42  & 20 & 0.021 & 0.021 \\
 17 $<$ V $<$ 18 & 23  & 8 & 0.025 & 0.031 \\
 18 $<$ V $<$ 19 & 85 & 38 & 0.029 & 0.037 \\
 19 $<$ V $<$ 20 & 41 & 20 & 0.040 & 0.042 \\
\hline
\end{tabular}
\caption{The standard deviation in the value of a single observation
of $V$ and $B-V$, as a function of $V$, evaluated from repeat
observations in the survey.  There are 215 repeat observations of 92
stars.\label{phot_sd}}
\end{table}

The external errors for the photometric measurements were determined
using the approach of Pearson \& Hartley (1976) (see also e.g. Kinman,
Suntzeff, and Kraft, 1994; hereafter KSK), which looks at the
range of values $R$ (i.e. the difference between the largest and
smallest values) in $N$ repeat observations of the same star.
The estimate of the external standard deviation, for a
single observation, is then given by $\sigma_{ext} = {R / k}$, where the
coefficient $k$ depends on $N$ (Table 22 of Pearson \& Hartley, 1976).
Assuming all stars in the same magnitude range have similar errors, we
can average the estimate of $\sigma$. For measurements of $n$ stars
the estimate of the external standard deviation is given by,

\begin{equation}
\sigma_{ext}=\frac{1}{n}\sum_{i=1,n}\frac{R_i}{k_i(N_i)}
\end{equation}

where $k_i(N_i)$ is the $k$ factor for star $i$ with $N_{i}$ repeat
observations. 

We used this approach to compute the errors on $V$ and $B-V$, in 5
magnitude bins. Note that the error on $B-V$ may be smaller than the
quadrature sum of the errors on $B$ and $V$, since the $BV$
observations were made quasi--simultaneously, and on closely the same
area of the chip, meaning that any systematic errors on the $B$ and
$V$ magnitudes may be correlated.  Before computing the errors we have
to be confident the sample does not contain variable stars, for
example RR Lyraes. Previous investigations of large samples of distant
A-type stars contain a significant number of variable stars,
particularly in the instability strip at $0.2\leq (B-V)_0 \leq 0.4$.
For example, KSK discovered 28 variable stars from a complete sample
of 213 A-type stars, and Wilhelm et al. 1999a discovered 56 out of a
complete sample of 1000.  These studies found very few variable stars
in the region $0.0\leq (B-V)_0 \leq 0.2$ (KSK find 3 stars, Wilhelm et
al. 1999a only 2). Accordingly, to compute the errors we limited our
analysis to this colour range. Over this colour range we have 92 stars
with repeat observations, with a total of 215 observations.  Table
\ref{phot_sd} provides the computed standard deviations, for a single
observation, for $V$ and $B-V$, in the 5 magnitude bins. The errors in
$V$ computed in this way are about $30\%$ larger than the random
errors, and indicate systematic errors contributing to the overall
error budget at a level comparable to the random errors. On the other
hand the errors in $B-V$ are only slightly larger than the random
errors.

We have adopted the values in Table \ref{phot_sd} as our error
estimates, reducing these values by $\sqrt{N}$, for $N$ repeat
observations.  Examination of Table \ref{phot_sd} reveals that, in
order to achieve standard errors in $B-V$ of $<0.03$ magnitudes, at
least 2 observations are required for stars fainter than $V=18$. This
was achieved for the majority, although not all, of our targets.

The photometric survey resulted in $BV$ observations of 280 stars.
The results are provided in Table \ref{phot_tabs}, which for each star
lists the name, the equatorial coordinates, the mean unextinguished
$V$ magnitude, and unreddened $(B-V)_0$ colour, and the uncertainties
for these two quantities. Also provided are the value of E($B-V$) used
for the extinction corrections, taken from Schlegel et al. (1998), and
$N$ the number of repeat observations. We discovered 3 stars with
colours $0.2\leq (B-V)_0 \leq 0.4$ that have large residuals
($\sigma_{(B-V)} \sim$ 0.1). These are candidate variable stars, and
have been excluded from the classification process. These three
suspected variables are marked with a `v'. The number of stars in each
field with CCD photometry is listed in Table \ref{observ_total}.

\begin{table*}   
\label{phot_tabs}
\begin{center}
\begin{tabular}{llllclccl}
\hline
\noalign{\smallskip}   
\multicolumn{1}{c}{Star} &   
\multicolumn{1}{c}{R.A. (J2000.0)} &     
\multicolumn{1}{c}{Dec. (J2000.0)} &   
\multicolumn{1}{c}{$V$} &   
\multicolumn{1}{c}{$(B-V)_0$} &
\multicolumn{1}{c}{$\sigma(V)$} & 
\multicolumn{1}{c}{$\sigma(B-V)$} &
\multicolumn{1}{c}{E$(B-V)$} &
\multicolumn{1}{c}{N} \\   
\noalign{\smallskip}   
\hline  
\noalign{\smallskip}  
CSGP-003        & 01 02  44.2 & -26 27  46.8 & 18.69 &  0.22 &  0.021 &   0.026 &   0.016 &  2 \\
CSGP-005        & 01 01  56.7 & -25 38  03.1 & 18.53 &  0.17 &  0.017 &   0.021 &   0.028 &  3 \\
CSGP-006        & 01 01  44.0 & -26 58  32.6 & 19.36 &  0.17 &  0.028 &   0.030 &   0.018 &  2 \\
CSGP-009        & 00 56  03.2 & -26 27  30.3 & 19.13 &  0.28 &  0.028 &   0.030 &   0.019 &  2 \\
CSGP-011        & 00 55  40.7 & -26 53  01.8 & 18.65 &  0.13 &  0.021 &   0.026 &   0.018 &  2 \\
CSGP-017        & 00 47  42.5 & -25 47  17.9 & 19.34 &  0.12 &  0.028 &   0.030 &   0.015 &  2 \\
CSGP-018        & 00 46  00.8 & -26 19  38.5 & 19.36 &  0.26 &  0.028 &   0.030 &   0.012 &  2 \\
CSGP-021        & 00 42  41.7 & -26 20  55.6 & 19.26 &  0.16 &  0.028 &   0.030 &   0.011 &  2 \\
CSGP-029        & 00 58  09.6 & -27 23  15.6 & 19.35 &  0.11 &  0.028 &   0.030 &   0.020 &  2 \\
\noalign{\smallskip}   									       
\hline 											       
\end{tabular}
\caption{Photometric data for the blue horizontal branch star
candidates, corrected for extinction.  Suspected variables are
labelled 'v'. Stars that were subsequently found to be quasars are
labelled 'q'. Stars in the complete sample have names prefixed `C',
otherwise the stars are prefixed 'I'. This table is presented in its
entirety in the electronic edition of Monthly Notices.  A portion is
shown here for guidance regarding its content and
form.\label{phot_tabs}}
\end{center}
\end{table*}

\begin{table}
 \centering
  \begin{tabular}{@{}lrrrrrrr}
   \hline \\[-12pt]
	Field name &
 \multicolumn{1}{c}{R.A.} & \multicolumn{1}{c}{Dec.} & \multicolumn{1}{c}{Nphot} & \multicolumn{1}{c}{Nspec} \\
	           & 
 \multicolumn{2}{c}{(J2000.0)} & &\\
	\hline \\[-12pt]
SGP  &     0 55  &  -27 47 & 22 & 24\\
SA94 &     2 53  &    0 12 & 34 & 17\\
F358 &     3 38  &  -34 50 & 44 & 8\\
F854 &    10 23  &   -0 15 & 75 & 24\\
F789 &    12 43  &   -5 16 & 63 & 27\\
MT   &    22 06  &  -18 39 & 42 & 56\\
\hline
\end{tabular}
\caption{The number of photometric and spectroscopic observations,
Nphot and Nspec made in each field. The survey contains photometric
measurements for 280 stars. There are 156 stars with spectroscopic
measurements. \label{observ_total}}
\end{table}

\section{Spectroscopy}

\subsection{Observations}

We obtained medium resolution optical spectra of 156 survey stars over
a total of 21 clear nights at the 4.2-m William Herschel Telescope
(WHT), La Palma, and the 3.9-m Anglo-Australian Telescope (AAT),
Siding Spring Observatory, Australia. The number of stars in each
field is provided in Table \ref{observ_total}. The dates of the
observations are listed in Table \ref{spectro_observe_tab}. The aim
was to restrict spectroscopy as far as possible to stars with
confirmed CCD colours in the range $0<(B-V)_0<0.2$, accurate to better
than 0.03 magnitudes, so that we could apply the {\em
$D_{0.15}$--Colour} classifier. However, because the photometric and
spectroscopic surveys proceeded simultaneously, this is not true for
all the stars with spectra. There are some stars without CCD
photometry at all, but it is still possible to classify them, using
the {\em Scale width--Shape} classifier.

In all observations the slit width was adjusted to be slightly greater
than the image FWHM.  Such an approach produces a close to optimal
balance between minimising loss of flux from the target and minimising
the signal from the sky background, while ensuring that the slit is
narrow enough that the centroid of the light from the star is known to
be accurately at the centre of the slit.  The latter consideration is
important in order to ensure that systematic errors in the radial
velocity determinations are minimised.  Observations in the MTF, SA94,
F854, and F789 fields were obtained with the WHT using the blue arm of
the ISIS spectrograph and an EEV12 2K CCD detector.  A 600 line
mm$^{-1}$ grating was used, resulting in a dispersion of 0.44 {\AA}
pixel$^{-1}$.  Typical FWHM resolutions were measured from unblended
comparison arc lines, and were found to be 3.5-4 pixels, or 1.5-1.8
{\AA}.  The spectral coverage was 3570 {\AA} to 5040 {\AA}, which
includes the H$\delta$, H$\gamma$ and H$\beta$ lines, and the Ca II K
3934 {\AA} line.  Observations in the SGP, F358 and MTF fields were
made at the AAT, using the RGO spectrograph with a 1024x1024 TEK CCD
chip and a 1200 line mm$^{-1}$ grating, resulting in a dispersion of
0.79 {\AA} pixel$^{-1}$, and a FWHM resolution of typically 4 pixels,
or $\sim 3.2${\AA}.  The spectral coverage of these observations was
3705 {\AA} to 4500 {\AA}, which includes the H$\delta$, H$\gamma$, and
Ca II K lines, but not the H$\beta$ line.

To achieve the minimum continuum $S/N$ ratio of 15 ${\mathrm\AA}^{-1}$,
to classify the stars (Paper I), entailed total integration times for
the targets within the range 800 -- 6000 sec, depending on the
brightness of the target and the observing conditions. The longer
integration times were broken into exposures of maximum length 1800
sec., in order to eliminate cosmic rays. In six cases the integration
was halted after the first exposure, when the spectrum revealed the
target to be a quasar.  The stars were observed as near culmination as
possible, with 85\% of spectra obtained at airmass $<$
1.3. Wavelength calibrations were obtained either before or after each
target exposure. This was sufficient, as instrument flexure over 1800
sec. was negligible.

Nightly observations were made of BHB radial velocity standard stars
selected from the Globular clusters M13, M15, M92 and NGC 288 (see
Table \ref{rvstar_spec_published} and references therein). We obtained
high $S/N$ observations every night, $S/N=40-120\,{\mathrm\AA}^{-1}$,
as templates for cross--correlation measurement of the radial
velocities of the targets. We also undertook an extensive set of
measurements at lower $S/N$, similar to that of the targets,
$S/N=15\,{\mathrm\AA}^{-1}$, as a means of quantifying the radial
velocity errors.  

\begin{table}
\centering
  \begin{tabular}{@{}lllll} \hline \\[-12pt] Telescope & Date & Fields
   \\ \hline \\[-12pt] 
AAT & 1993\,Sep 12,14,16,17 & MTF, SGP \\
AAT & 1994\,Sep 7-9 & MTF, SGP \\ 
AAT & 2000\,Sep 2-4 & SGP, F358, MTF \\
WHT & 2000\,Aug 23-28 & SA94, MTF \\ 
WHT & 2001\,Feb 27-Mar 3 & F854, F789 \\

\hline
\end{tabular}
\vspace{5mm}
\caption{Dates of spectroscopic observations, and fields
observed. \label{spectro_observe_tab}}
\end{table}

We used standard data reduction routines available in {\tt IRAF} for
bias and flat field correction, and removal of cosmic rays.  We then
extracted sky-subtracted variance-weighted one--dimensional spectra
using a wide aperture. The wavelength-calibration arc spectra were
extracted using the corresponding traced target-star apertures. A
polynomial of degree three\footnote{in IRAF parlance this is a
polynomial of order four} was fit to the $30-45$ arc lines identified in
each spectrum.

The measurement of the Balmer line widths and shapes followed the
procedures of Paper I exactly, averaging the results for the $H\delta$
and $H\gamma$ lines.  After dividing by the polynomial fit to the
continuum, we used our custom Sersic--profile fitting software to
measure the relevant parameters: the line-width at $15\%$ depth
$D_{0.15}$, the scale width $b$, and the shape index $c$.  The routine
measures deconvolved quantities, and requires the resolution of each
spectrum as input.  Measures of the FWHM from unblended comparison arc
lines are available for all the object spectra.
Comparing individual measurements to those with a FWHM fixed at the
mean value gives absolute differences of at most 3\% in any parameter.
This suggests that the Balmer line measurements are relatively
insensitive to uncertainties in the estimation of the resolution.

The results of the measurements of the Balmer lines are provided in
Table \ref{spec_tabs}.  Column (1) gives the name of the star, and
columns (2) to (4) list, respectively, the parameters $D_{0.15}$, $b$,
and $c$.  The errors on the parameters $b$ and $c$ are provided in
columns ((5) to (7), in the form of $A$ and $B$, the semi-major and
semi-minor axes of the error ellipse in the $b-c$ plane, and $\theta$
the orientation of the semi-major axis, measured anti-clockwise from
the $b$-axis.  Here the error corresponds to the $68\%$ confidence
interval for each axis in isolation (see Paper I for further details).
We also fit the profiles of the Balmer lines of the radial velocity
standards (\S4.3), and the results are provided in Table
\ref{rvstar_spec_published_res}.

\begin{table*}
\begin{center}
\end{center}
\begin{center}
 \begin{tabular}{lrccccccrcc}
\hline
\noalign{\smallskip}   
\multicolumn{1}{c}{Star} &   
\multicolumn{1}{c}{$D_{0.15}(\gamma\delta)$} &   
\multicolumn{1}{c}{$b(\gamma\delta)$} &    
\multicolumn{1}{c}{$c(\gamma\delta)$} &
\multicolumn{1}{c}{$A$} &  
\multicolumn{1}{c}{$B$} &  
\multicolumn{1}{c}{$\theta$} &  
\multicolumn{1}{c}{[Fe/H]} &   
\multicolumn{1}{c}{V$_{\odot}$} &     
\multicolumn{1}{c}{$d_{\odot}$} &   
\multicolumn{1}{c}{Classif.} \\
\multicolumn{1}{c}{} &  
\multicolumn{1}{c}{[\AA]} &
\multicolumn{1}{c}{[\AA]} &   
\multicolumn{1}{c}{} &   
\multicolumn{1}{c}{} &      
\multicolumn{1}{c}{} &    
\multicolumn{1}{c}{} &
\multicolumn{1}{c}{} &  
\multicolumn{1}{c}{[km s$^{-1}$]} &  
\multicolumn{1}{c}{[kpc]} &  
\multicolumn{1}{c}{} \\
\noalign{\smallskip}   
\hline  
\noalign{\smallskip} 
CSGP-003            &  31.41 $\pm$   0.96 &    7.25 &   0.69 &   0.39 &  0.028 &  1.519 &   -1.4 $\pm$   0.2 &  -46.24 $\pm$  10.2 &   14.5 $\pm$   2.0 &    A/BS\\ 
CSGP-005            &  35.85 $\pm$   1.39 &    9.25 &   0.78 &   0.55 &  0.051 &  1.520 &   -1.4 $\pm$   0.2 &  124.13 $\pm$  14.2 &   14.6 $\pm$   1.3 &    A/BS\\ 
CSGP-006            &  35.45 $\pm$   1.05 &    8.40 &   0.69 &   0.41 &  0.029 &  1.524 &   -1.6 $\pm$   0.2 & -103.82 $\pm$  10.6 &   20.6 $\pm$   1.4 &    A/BS\\ 
CSGP-009            &  21.91 $\pm$   1.15 &    5.18 &   0.67 &   0.38 &  0.084 &  1.481 &   -0.9 $\pm$   0.3 &  -65.80 $\pm$  13.5 &   45.2 $\pm$   3.3 &     BHB\\ 
CSGP-011            &  37.86 $\pm$   1.07 &    9.65 &   0.79 &   0.44 &  0.034 &  1.525 &   -1.3 $\pm$   0.2 &  -41.81 $\pm$  13.2 &   17.0 $\pm$   1.6 &    A/BS\\ 
CSGP-017            &  35.13 $\pm$   1.49 &    9.87 &   0.93 &   0.61 &  0.068 &  1.520 &   -1.9 $\pm$   0.4 &   27.25 $\pm$  16.1 &   53.5 $\pm$   4.6 &     BHB\\ 
CSGP-018            &  23.92 $\pm$   1.71 &    4.97 &   0.61 &   0.64 &  0.049 &  1.509 &   -1.3 $\pm$   0.2 &   89.22 $\pm$  16.0 &   18.8 $\pm$   1.8 &    A/BS\\ 
CSGP-021            &  32.99 $\pm$   1.58 &    8.17 &   0.75 &   0.65 &  0.052 &  1.518 &   -2.0 $\pm$   0.3 &  -89.94 $\pm$  15.2 &   52.0 $\pm$   4.4 &     BHB\\ 
CSGP-029            &  31.08 $\pm$   0.77 &    7.27 &   0.70 &   0.31 &  0.024 &  1.518 &   -1.1 $\pm$   0.2 &  -68.40 $\pm$  10.1 &   50.7 $\pm$   4.5 &     BHB\\ 
CSGP-030            &  34.77 $\pm$   1.20 &    8.87 &   0.78 &   0.47 &  0.043 &  1.522 &   -1.8 $\pm$   0.3 &  -45.18 $\pm$  13.4 &   16.8 $\pm$   1.8 &    A/BS\\ 
CSGP-034            &  28.89 $\pm$   1.20 &    7.09 &   0.75 &   0.45 &  0.044 &  1.517 &   -1.2 $\pm$   0.3 &  -41.25 $\pm$  13.6 &   51.2 $\pm$   4.0 &     BHB\\ 
\noalign{\smallskip}   
\hline 
\label{tab:sgp}
\end{tabular}
\caption{Spectroscopic data for the horizontal branch star
candidates. This table is presented in its entirety in the electronic
edition of Monthly Notices. A portion is shown here for guidance
regarding its content and form.\label{spec_tabs}}
\end{center}
\end{table*}

\subsection{Metallicities from Ca II K lines} 

The Ca II K line is the strongest measurable metal line present in
the wavelength range covered by the spectra, and the only useful line
in moderate resolution blue spectra for measuring metallicity. With
the exception of the Am and Ap stars, the Ca II K line equivalent
width EW, in conjunction with the $(B-V)_0$ colour, provides a
moderately accurate measure of the metallicity of the star (e.g.  Pier
1983, Beers et al. 1992, KSK).

\begin{figure}
\rotatebox{0}{
\centering{
\scalebox{0.30}{
\includegraphics*[-50,140][700,700]{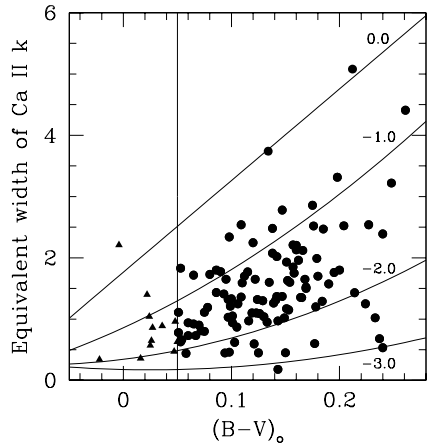}
}}}

\caption{Ca II K line EW measurements for the survey stars. The curves
of constant metallicity for [Fe/H] = -1.0, -2.0 and -3.0 were computed
by Wilhelm et al. (1999a) and the line for [Fe/H] = 0.0 is taken from
Paper I. Stars to the left of the vertical line at $(B-V)_0$ $=$ 0.05
have indeterminate metallicities (filled triangles). Measurements for
the remaining stars are shown as filled circles.\label{cak_plot}}
\end{figure}

In Paper I we described our procedures for measuring the Ca II K line
EW, and for estimating the metallicity. We follow the procedures of
Paper I here, exactly. The method is illustrated in
Fig. \ref{cak_plot}, which plots EW$_{Ca}$ of the survey stars versus
$(B-V)_0$. As explained in Paper I, the metallicity is measured by
interpolating between the plotted curves of fixed metallicity. In
Paper I we calibrated the intrinsic accuracy of this method, using
high $S/N$ spectra (i.e. negligible error in EW$_{Ca}$), finding a
dispersion of 0.3dex in the measured metallicities relative to accurate
literature values from high-resolution spectra. We have confirmed this
value by measurements of the BHB radial velocity standards (\S4.3) used in
this paper. For the survey stars, this intrinsic scatter of 0.3dex is
added in quadrature to the uncertainty in the metallicity associated
with the accuracy of the measured EW. Because the curves crowd
together for bluer colours, we have not attempted to measure the
metallicities of stars with colours $(B-V)_0<0.05$.

The measured metallicities and their random errors (only) are listed
in column (8) of Table \ref{spec_tabs}.  The majority of the survey
stars are metal poor, with a mean metallicity [Fe/H]$=-1.7$ with
dispersion 0.46, and all but two stars have measured metallicities in
the range $-3.0 <$ [Fe/H] $< -0.5$.  The two metal--rich stars are
CF789-093 and CMTF-058, which both have measured values close to
solar. There are 13 stars bluer than $(B-V)_0$ $=$ 0.05 (shown as
filled triangles), for which we cannot estimate the metallicity
reliably.

\subsection{Radial velocity determinations} 

Radial velocities for the stars were measured by cross-correlation of
the spectra with our high $S/N$ spectra of BHB template stars of known
heliocentric radial velocity. Relevant details of the template stars,
taken from the literature, are provided in Table
\ref{rvstar_spec_published}. Here the quoted metallicities are average
values for each globular cluster, and are based on high--resolution
spectra.  We used the XCSAO package in {\tt IRAF}, which uses Fourier
methods, based on the methodology set out by Tonry \& Davis
(1979). After continuum subtraction, and transformation to a log
wavelength scale, the spectra are filtered in Fourier space, to remove
low frequency continuum variations and high frequency noise, and then
apodized, before cross-correlation. The peak of the cross-correlation
function was identified in each spectrum and fit with a parabola. The
position of the peak measures the difference in the radial velocities,
and the measured radial velocity was then corrected to the
heliocentric value. We checked that using different functional fits to
the cross-correlation peak produced very similar results.

The width of the cross-correlation peak provides an estimate of the
uncertainty in the radial velocity. These random errors were typically
7 km s$^{-1}$ for our survey stars. We then measured the scatter in
repeat measurements of the template stars, over different nights,
which was 10 km s$^{-1}$, significantly larger than the random errors
for these high S/N spectra. This scatter is plausibly due to the star
being imperfectly centred in the slit, as confirmed by measurements
where we deliberately offset a template star to the edge of the
slit. Accordingly we have combined this error in quadrature with the
random errors. The result is a typical accuracy of 12 km s$^{-1}$ for
the measured radial velocity of a survey star.  The scatter in repeat
measures of the radial velocity of a number of survey stars, as well
as template stars observed at lower $S/N$, is consistent with the
errors computed in this manner.
 
\begin{figure}
\rotatebox{0}{
\centering{
\scalebox{0.3}{
\includegraphics*[-50,140][700,700]{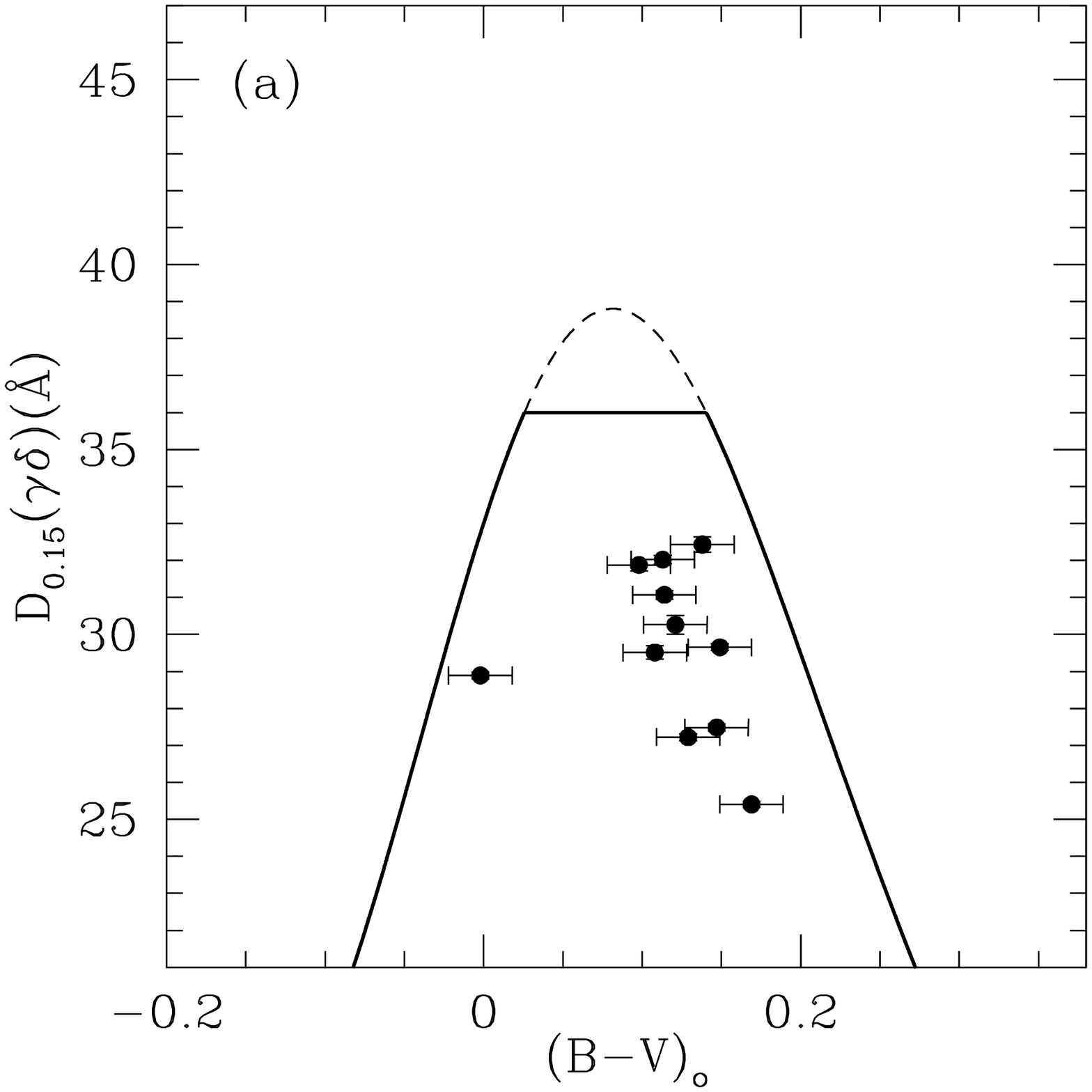}
}}}
%
\rotatebox{0}{
\centering{
\scalebox{0.3}{
\includegraphics*[-50,140][700,700]{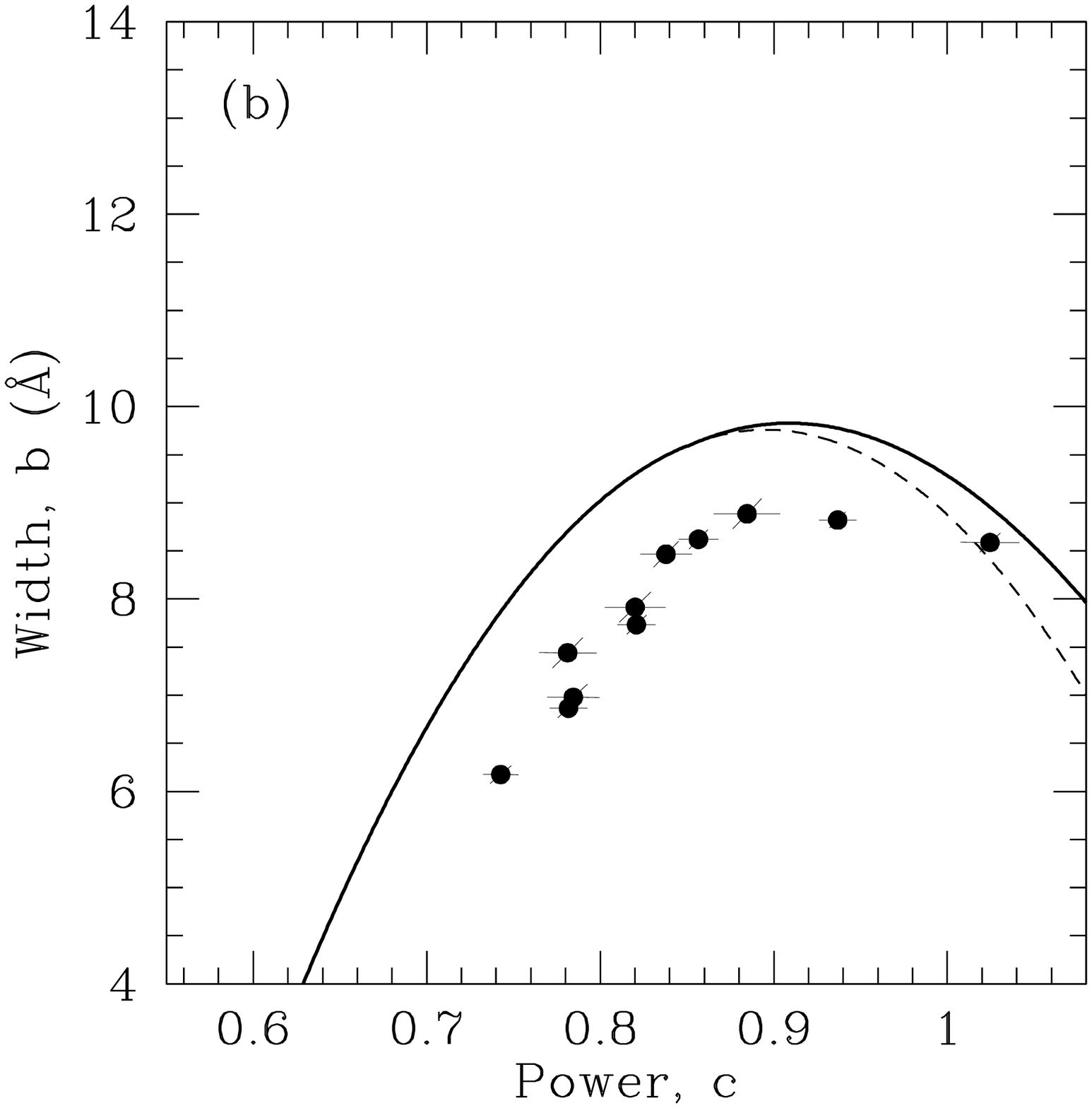}
}}}
\caption{Classification of 11 BHB radial velocity stars from Table
\ref{rvstar_spec_published_res} by (a) the {\em $D_{0.15}$--Colour}
method and (b) the {\em Scale width--Shape} method. The classification
boundaries determined in Paper I are shown as a dashed line. For the
{\em $D_{0.15}$--Colour} method we have imposed an upper limit of
$D_{0.15}$ = 36{\AA} shown as a solid line. One star, M92 IV-17, is
misclassified in the {\em Scale width--Shape} plot and an adjusted
form for the classification boundary has been adopted for values of $c
> 0.85$ (solid line).\label{width_rv}}

\end{figure}
\begin{table*}   
\begin{flushleft}
\begin{center}
\begin{tabular}{lrrrrrrrrrrrrrrrr}
\hline
\noalign{\smallskip}   
\multicolumn{1}{c}{Star} &   
\multicolumn{1}{c}{R.A. (2000)} &     
\multicolumn{1}{c}{Dec.} &   
\multicolumn{1}{c}{V} &   
\multicolumn{1}{c}{B-V} &
\multicolumn{1}{c}{E(B-V)} &
\multicolumn{1}{c}{V$_{\odot}$} &
\multicolumn{1}{c}{[Fe/H]} & 
\multicolumn{1}{c}{Source} \\  
\multicolumn{1}{c}{} &  
\multicolumn{1}{c}{} &  
\multicolumn{1}{c}{} &  
\multicolumn{1}{c}{} &  
\multicolumn{1}{c}{} & 
\multicolumn{1}{c}{} &  
\multicolumn{1}{c}{[km s$^{-1}$]} &
\multicolumn{1}{c}{} &  
\multicolumn{1}{c}{} \\
\noalign{\smallskip}   
\hline  
\noalign{\smallskip}
NGC 288-431 &  0 52 37.9  & -26 36 34.2  &  15.22  &  0.15  & 0.012 &  -45.1$\pm$  1.7 & -1.24 & (2), (3), (5) \\
NGC 288-421 &  0 52 38.3  & -26 37 03.0  &  15.26  &  0.12  & 0.012 &  -39.2 $\pm$ 1.7 & -1.24 & (2), (3), (5) \\
NGC 288-418 &  0 52 39.2  & -26 37 28.0  &  15.22  &  0.11  & 0.012 &  -39.0 $\pm$ 2.9 & -1.24 & (2),(3), (5) \\
M5 II-78    & 15 18 27.6  &  02 07 25.8  &  15.02  &  0.16  & 0.039 &  +42.2 $\pm$ 1.1 & -1.33 & (1), (5) \\
M13 III-58  & 16 41 37.5  &  36 25 17.2  &  15.07  &  0.13  & 0.016 & -247.4 $\pm$ 2.2 & -1.56 & (1), (5) \\
M13 III-26  & 16 41 30.2  &  36 26 13.2  &  14.99  &  0.13  & 0.017 & -254.1 $\pm$ 2.0 & -1.56 &  (1), (5) \\
M92 IV-17   & 17 17 00.9  &  43 13 04.2  &  15.50  &  0.02  & 0.022 & -127.6 $\pm$ 2.0 & -2.29 &	(4), (5) \\
M92 IV-27   & 17 17 06.8  &  43 12 23.9  &  15.19  &  0.17  & 0.021 & -115.7 $\pm$ 2.0 & -2.29 &	(4), (5) \\
M92 XII-9   & 17 17 30.1  &  43 06 08.3  &  15.09  &  0.15  & 0.021 & -127.7 $\pm$ 2.0 & -2.29 &(4), (5) \\
M92 XII-1   & 17 17 45.1  &  43 05 59.0  &  15.11  &  0.19  & 0.021 & -127.6 $\pm$ 2.0 & -2.29 &	(4), (5) \\
M15 IV-44   & 21 30 02.8  &  12 08 36.7  &  15.79  &  0.25  & 0.103 &-108.5 $\pm$ 2.1 & -2.22 &  (1), (5) \\
\noalign{\smallskip}
\hline 
\end{tabular}
\caption{Published spectroscopic and photometric data for selected
globular cluster horizontal branch stars.\label{rvstar_spec_published}}
\end{center}
Sources: (1) Peterson, 1983 (M3, M5 and M13); (2) Peterson, 1985 (NGC 288); (3) Olszewski, Harris \& Canterna 1984 (CM photometry for NGC 288); (4) Cohen \& McCarthy 1997 (M92) and (5) Harris 1996.
\end{flushleft}
\end{table*}
\begin{table*}   
\begin{flushleft}
\begin{center}
\begin{tabular}{lrlrrrrrr}
\hline
\noalign{\smallskip}   
\multicolumn{1}{c}{Star} & 
\multicolumn{1}{c}{$D_{0.15}(\gamma\delta)$} &
\multicolumn{1}{c}{$b(\gamma\delta)$} & 
\multicolumn{1}{c}{$c(\gamma\delta)$} &
\multicolumn{1}{c}{$A$} &  
\multicolumn{1}{c}{$B$} &  
\multicolumn{1}{c}{$\theta$} &  
\multicolumn{1}{c}{$d_{\odot}$} &   
\multicolumn{1}{c}{N} \\
\multicolumn{1}{c}{} &  
\multicolumn{1}{c}{[\AA]} &      
\multicolumn{1}{c}{[\AA]} &      
\multicolumn{1}{c}{} &    
\multicolumn{1}{c}{} &
\multicolumn{1}{c}{} &  
\multicolumn{1}{c}{} &  
\multicolumn{1}{c}{[kpc]} &  
\multicolumn{1}{c}{} \\
\noalign{\smallskip}   
\hline  
\noalign{\smallskip}   
NGC 288-431  & 32.43 $\pm$    0.21  &  7.4    $\pm$     0.3   &     8.88     &  0.89    &       0.16  &   0.019   &  1.518   &4\\
NGC 288-421  & 29.51 $\pm$    0.18  &  7.3    $\pm$     0.2   &     7.44     &  0.78    &       0.16  &   0.017   &  1.515   &5\\
NGC 288-418  & 31.88 $\pm$    0.15  &  7.2    $\pm$     0.2   &     8.47     &  0.84    &       0.13  &   0.015   &  1.517   &5\\
M5 II-78     & 30.26 $\pm$    0.25  &  6.8    $\pm$     0.3   &     7.91     &  0.82    &       0.16  &   0.018   &  1.513   &3\\
M13 III-58   & 31.07 $\pm$    0.12  &  7.0    $\pm$     0.5   &     8.82     &  0.94    &       0.08  &   0.011   &  1.514   &3\\
M13 III-26   & 32.02 $\pm$    0.12  &  6.8    $\pm$     0.2   &     8.62     &  0.86    &       0.10  &   0.011   &  1.517   &5\\
M92 IV-17    & 28.89 $\pm$    0.09  &  8.1    $\pm$     0.2   &     8.59     &  1.02    &       0.10  &   0.017   &  1.511   &11\\
M92 IV-27    & 29.65 $\pm$    0.09  &  8.2    $\pm$     0.2   &     7.73     &  0.82    &       0.10  &   0.011   &  1.512   &10\\
M92 XII-9    & 27.22 $\pm$    0.09  &  7.7    $\pm$     0.2   &     6.87     &  0.78    &       0.10  &   0.011   &  1.509   &10\\
M92 XII-1    & 25.40 $\pm$    0.08  &  8.0    $\pm$     0.2   &     6.18     &  0.74    &       0.09  &   0.010   &  1.506   &11\\
M15 IV-44    & 27.48 $\pm$    0.10  &  10.8   $\pm$     0.2   &     6.98     &  0.78    &       0.14  &   0.015   &  1.511   &16\\
\noalign{\smallskip}   
\hline 
\end{tabular}
\caption{Spectroscopic results for selected globular cluster horizontal branch stars.\label{rvstar_spec_published_res}} 
\end{center}
\end{flushleft}
\end{table*}
\section{Distance measurements}
The absolute magnitude of a BHB star $M_V(BHB)$ depends on both colour
(i.e. temperature) and metallicity.  By assuming all stars in a
globular cluster have the same metallicity, the colour dependence may
be determined (independent of knowledge of the distance to the
cluster) by plotting the difference between the apparent magnitude of
the stars along the horizontal branch, and the apparent magnitude of
the RR Lyraes in the cluster. Stacking the results for several
clusters, Preston et al. (1991) derived a cubic relation in $(B-V)_0$,
which we have adopted. To determined the absolute magnitude of a BHB
star of given metallicity, we then need to know the dependence on
metallicity of the absolute magnitude of RR Lyrae stars, parameterised
as,
\begin{equation}  
M_V(RR) = \alpha + \beta[\rm{Fe/H}].
\label{equ:mv}
\end{equation} 
The exact form of Equation \ref{equ:mv} remains controversial. We
refer the interested reader to a recent review of this subject by
Cacciari \& Clementini 2003. We use recent observations by Clementini et
al. (2003) that employ $\sim 100$ RR Lyrae stars located in the Large
Magellanic Cloud to derive $\beta=0.214 \pm 0.047$ for the slope of
this relation. We fix the zero point by reference to the measurement
by Gould \& Popowski (1998) of the absolute magnitude of RR Lyrae
stars, $M_V(RR)$ = 0.77 $\pm$ 0.13 mag at [Fe/H] = -1.60, using
statistical parallaxes derived from {\it Hipparcos} observations. The
final expression for the absolute magnitude of BHB stars is then
\begin{eqnarray}  
M_V(BHB) &=& 1.552+0.214[\rm{Fe/H}]-4.423(B-V)_0 \nonumber\\
& & + 17.74(B-V)^2_0-35.73(B-V)^3_0. 
\label{ch4:abs_mag}
\end{eqnarray}
For typical values $(B-V)_0\sim0.1$, [Fe/H]$\sim-1.7$, this gives
$M_V\sim0.9$. Distances and associated errors are then derived using
the unextinguished apparent magnitudes, $V$, and the corresponding
photometric error.  We average repeat observations using inverse
variance weights, then add the error from the systematic error on the
metallicity. The result produces distance errors of 2-6\% for the
standard stars and 6-10\% for our survey objects. Distances computed
in this manner are provided in column (10) of Table
\ref{spec_tabs}. For the small number of stars with $(B-V)_0<0.05$,
for which we are unable to measure the metallicity, we adopted the
mean value. For interest we also computed distances in this way for
the radial velocity standards, using the accurate literature
metallicities (Table \ref{rvstar_spec_published}), and the results are
provided in Table \ref{rvstar_spec_published_res}.

Less is known about the absolute magnitudes of blue stragglers and so
their distances are much less accurate than those of the BHB stars. We
have adopted the following relation derived by KSK from data for blue
stragglers in globular clusters published by Sarajedini (1993):

\begin{equation}  
M_V(BS) = 1.32 + 4.05(B-V)_0 - 0.45[\rm{Fe/H}]. 
\end{equation} 

Nevertheless, recent work by Preston \& Sneden (2000) suggests that
field blue straggler stars may have entirely different formation
mechanisms from those observed in clusters, owing their origin to mass
transfer (McCrea 1964).  The transfer of mass in a close binary system
created by Roche-lobe overflow during red giant evolution can produce
a substantial increase in the main-sequence lifetime of the star.  It
is certainly possible that we are dealing with a mixture of different
types of stars, with each population characterised by distinct
absolute magnitude ranges and hence distances. We note also that the Preston \&
Sneden (2000) stars  are on average redder than the stars dicussed in
this paper and may not be completely identical. Notwithstanding these
caveats, we used the KSK--derived $M_V$ to calculate distances for
these objects, and stress that the quoted uncertainties reflect only
measurement errors.  We note, however, that Yanny et al. (2000) found
absolute magnitudes of blue stragglers in the halo that were some 2
magnitudes fainter than the BHB stars, entirely consistent with the above
estimates.

\begin{figure*}
\begin{minipage}{133mm}
\epsfig{figure=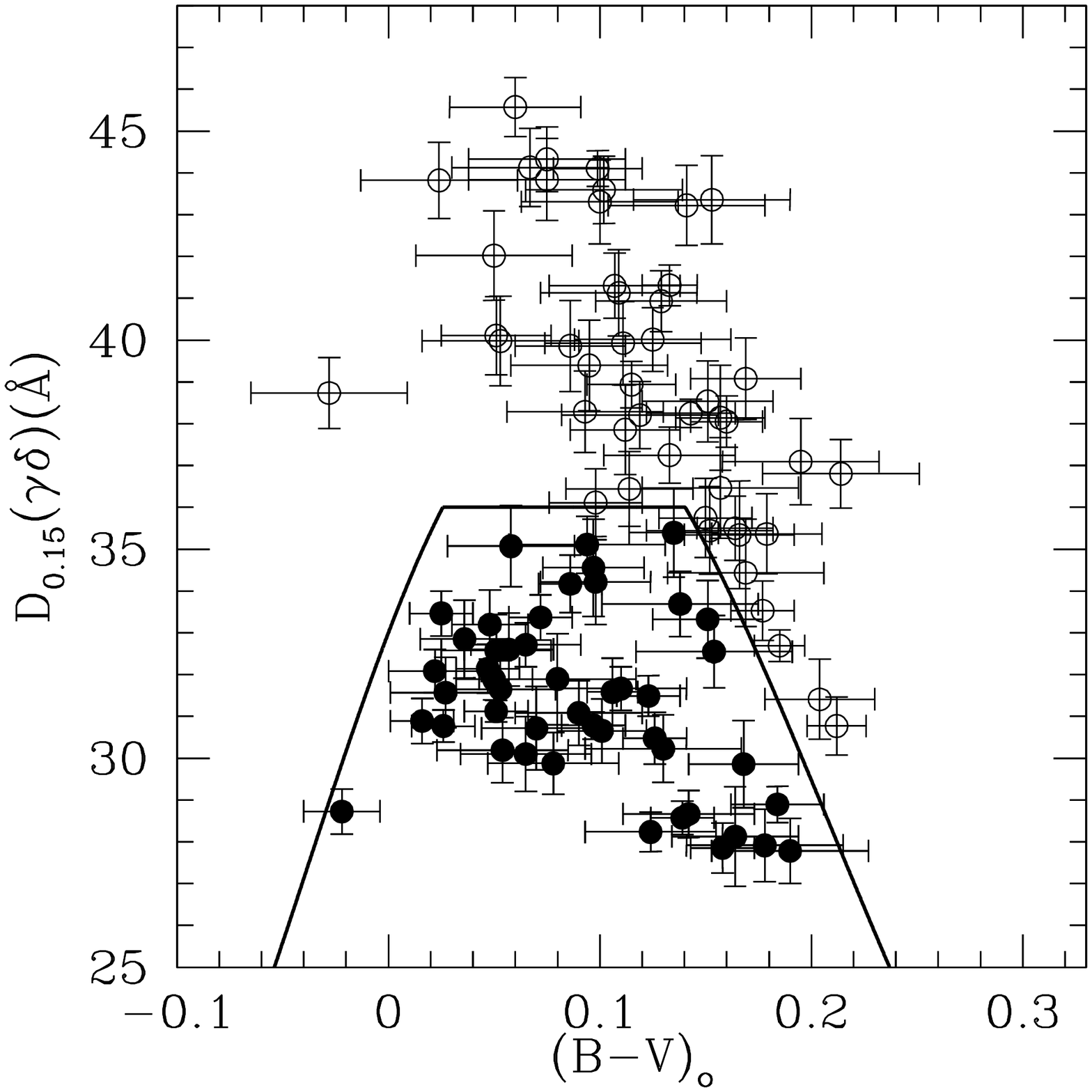,height=66mm,width=66mm}
\epsfig{figure=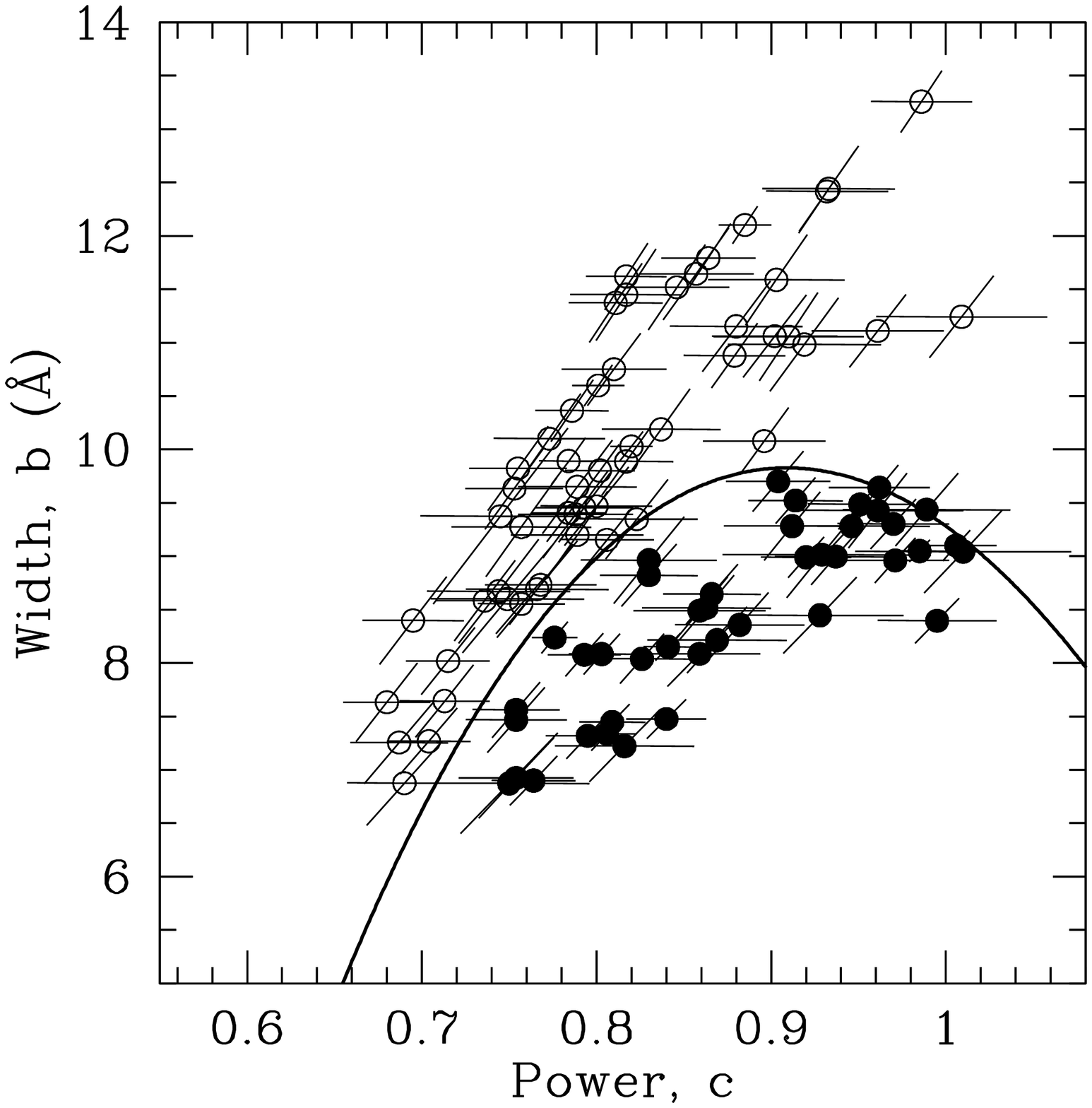,height=66mm,width=66mm}
\epsfig{figure=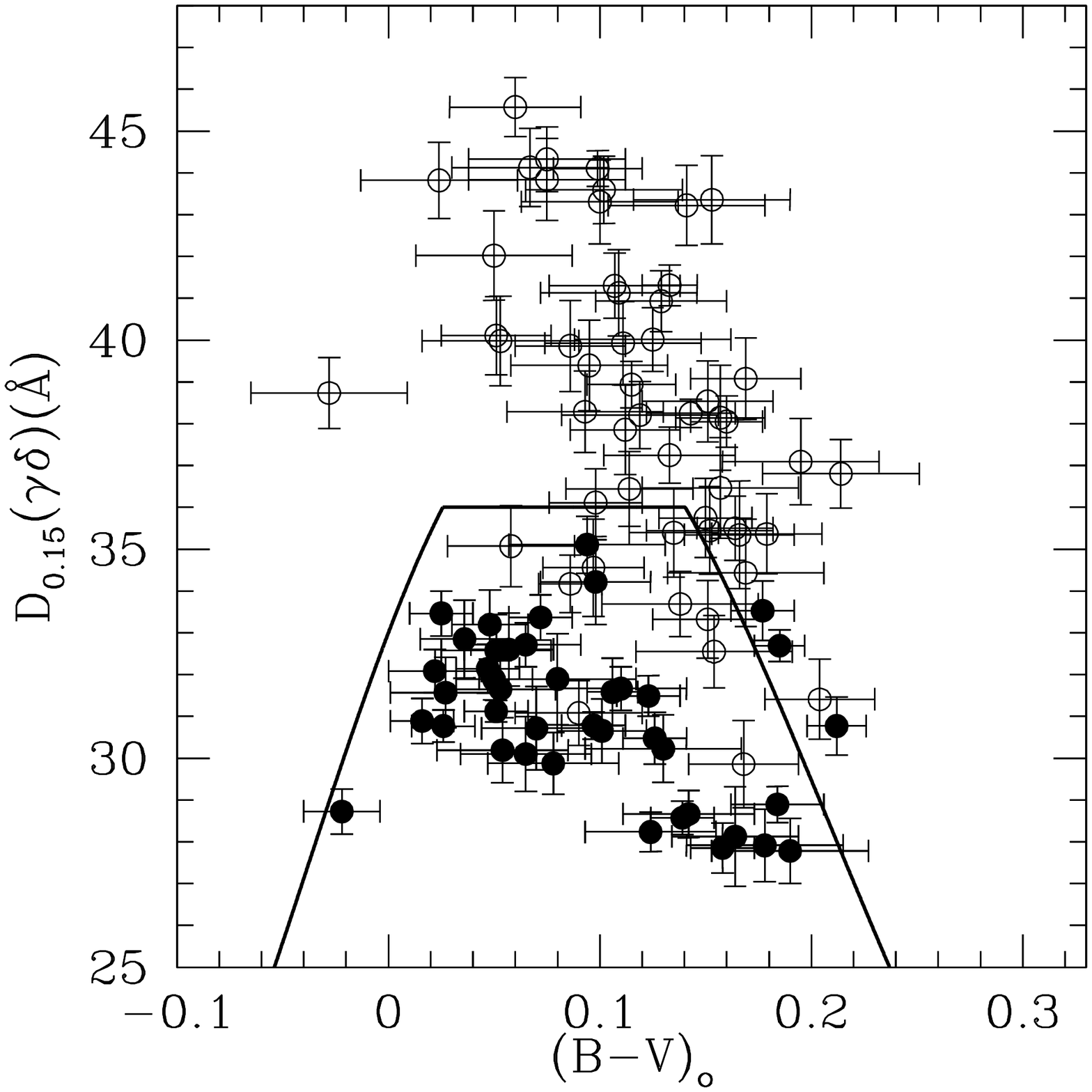,height=66mm,width=66mm}
\epsfig{figure=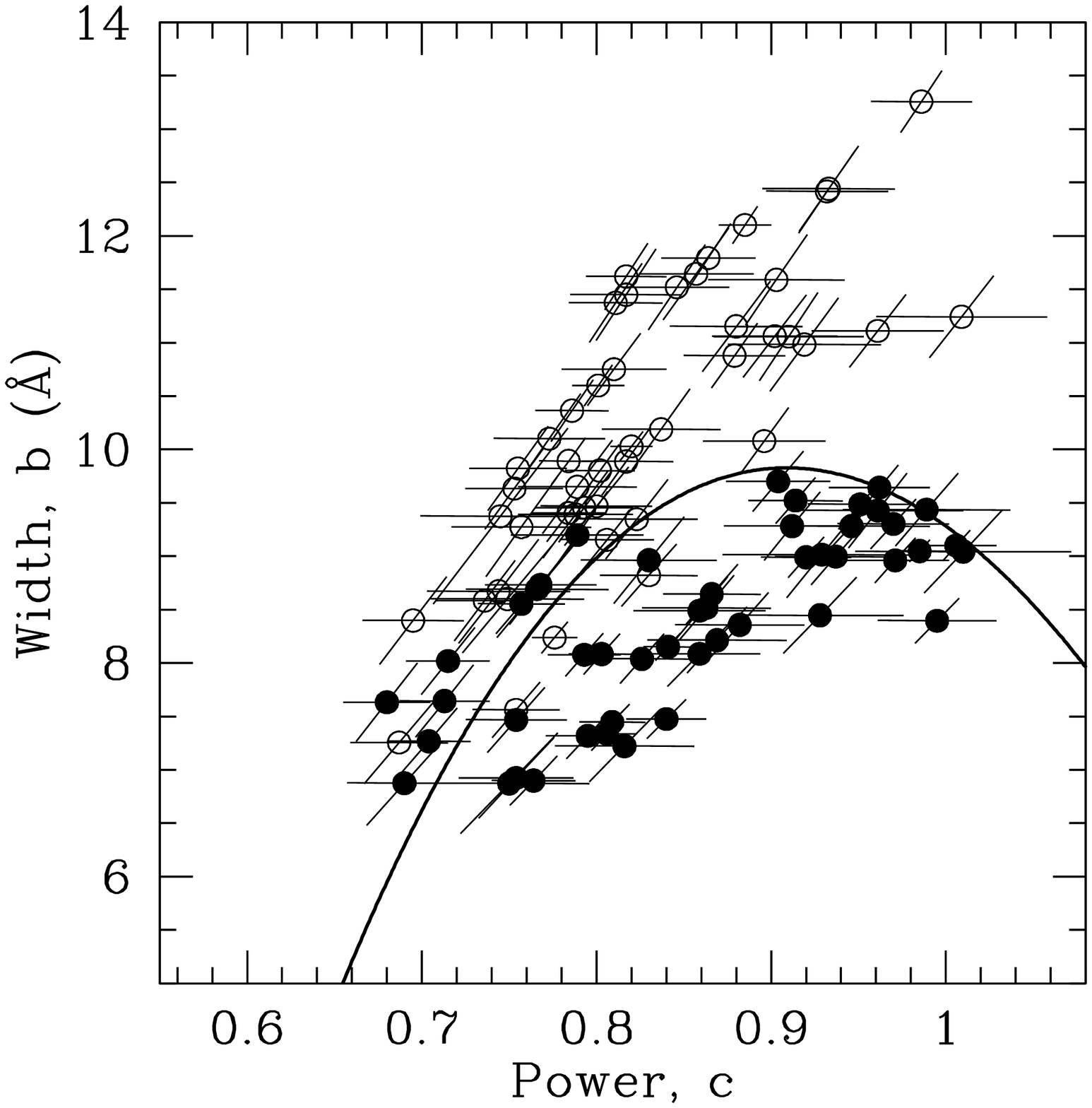,height=66mm,width=66mm}
\end{minipage}

\caption{Classification of the 94 programme stars in the photometric
sample (\S6.2), using the {\em $D_{0.15}$--Colour} (left) and the {\em
Scale width--Shape} (right) classification methods. The solid curves
are the classification boundaries explained in the text.  For the {\it
upper plots} filled circles are stars classified BHB in that plot,
i.e. the stars below the classification boundary.  In the {\em
$D_{0.15}$--Colour} plot there are 47 stars below the boundary and 47
above it. In the {\em Scale width--Shape} plot there are 41 stars
below the boundary and 53 above it. For the {\it lower plots}, in each
case filled circles mark stars classified BHB by the other
classification method.
\label{class_cw_sw}}
\end{figure*}

\section{Classification}\label{sec:comp}

\subsection{The BHB standard stars}

Before turning our attention to the survey objects, we first examine
how well our classifications methods work on the radial velocity
standards observed in M13, M15, M92 and NGC 288, which are all known
BHB stars.  Table \ref{rvstar_spec_published_res} lists the relevant
parameters measured from the spectra. These are $D_{0.15}$, $b$, $c$,
and their errors.

Figure \ref{width_rv} (a) plots $D_{0.15}(\gamma\delta)$ (average for
the H$\gamma$ and H$\delta$ lines) versus $(B-V)_0$ for the radial
velocity standards. The dashed line marks the classification boundary
determined in Paper I, which we here revise slightly to the solid
line, by imposing an upper limit $D_{0.15}=36$\AA, as explained below.
The dashed line is defined by the equation $D_{0.15} =
7918.2(B-V)_0^4 - 2409.4(B-V)_0^3 - 549.5(B-V)_0^2 + 125.2(B-V)_0 + 33.0$.
The plot shows that all 11 radial velocity standards are
correctly classified as BHB stars (using either curve). 

Brown et al. (2003) have applied our classification methods to a
sample of halo A-type stars. Their Fig. 6c, suggests that our
classification curve may be a little high, near $(B-V)_0=0.1$
(although note that their plot uses estimated rather than measured
$(B-V)_0$ colours). Referring to Fig. 5 Paper I, where the
classification curve was defined, the largest measured value for any
BHB star in the KSK sample is $D_{0.15}=35.1$\AA. Since
this parameter is measured to an accuracy of $\sim1$\AA\, for our
programme stars, by imposing an upper limit $D_{0.15}=36$\AA, only a
very small number of BHB stars will be lost from the sample, while
contamination by blue stragglers (already small) should be
significantly reduced.
 
Figure \ref{width_rv} (b) plots $b(\gamma\delta)$ versus
$c(\gamma\delta)$ for the radial velocity standards. The dashed line
marks the classification boundary determined in Paper I, which we
again here revise slightly, to the solid line. In this case, 10 of the
11 standard stars lie below the dashed line, and are correctly
identified as BHB stars.  The exception, M92 IV-17, has a value of $c
= 1.02$. This is larger than any of the values used in the original
definition of the classification boundary i.e. the dashed
classification line is an extrapolation in this region, and it appears
it may lie too low. Further evidence that our classification curve is
low in this region comes from the same measurements by Brown et
al. (2003), referred to above. In their Fig. 6d, at large $c>1.0$, two
confirmed BHB stars lie above the curve. Accordingly we have revised
the classification curve slightly upward in this region. The revised
classification boundary, shown by the solid curve, is given by the
expression $b = 25.7c^3 - 138.191c^2 + 187.5c -65.8$.

With this minor revisions we are now in a position to classify our
programme stars into categories BHB or A/BS. 

\subsection{The survey stars}\label{sec:class}
In Paper I we found halo BHB stars can be separated reliably from halo
blue stragglers provided the spectra have a $S/N> 15
{\mathrm\AA}^{-1}$ and EW H$\gamma>13$\AA\, (equivalent to the colour
range $0 \leq (B-V)_0 \leq 0.2$). There are 142 such stars in this
survey. This sample includes 94 stars which also have $B-V$ colours
accurate to $\le$ 0.03 magnitudes. We call this sub-sample the
`photometric sample', and use it to compare the classifications
between the two methods. The remaining 48 stars contain photometric
errors greater than 0.03 magnitudes and so are only classified using
the {\em Scale width--Shape} classification method.

Figure \ref{class_cw_sw} (upper) plots the 94 stars in the photometric
sample for the {\em $D_{0.15}$--Colour} and {\em Scale width--Shape}
classification methods. For the {\em $D_{0.15}$--Colour} plot (left)
there are 47 stars below the classification boundary (probable BHB
stars, plotted solid, hereafter sample A), and 47 stars above the
classification boundary (probable blue stragglers, plotted open). In
the {\em Scale width--Shape} plot (right) there are 41 stars
classified BHB (hereafter sample B) and 53 classified A/BS. Given the
predicted high completeness and low contamination of both methods
(Paper I) we would expect close agreement between the two
methods. Furthermore the two classifications are not completely
independent, since the parameters $D_{0.15},$ $b,$ and $c$ are related
through the equation $D_{0.15}=2b\mathrm{ln}(0.83/0.15)^{1/c}$.

The agreement between the two methods is indicated in the lower
plot. On the LHS, the 41 stars of sample B are plotted solid. On the
RHS, the 47 stars of sample A are plotted solid. The number below the
line plotted solid in each of these plots, i.e. the number of stars
classified BHB by both methods, is 38.  Over the range $0.8<c<1.05$
(lower RH plot) the agreement is essentially perfect, with only one
star in sample B (the single open symbol below the line) not in sample
A, and all the stars in sample A also in sample B (i.e. all the black
symbols are below the line). The agreement is less good for the region
$c<0.8$ (corresponding approximately to $(B-V)_0>0.14$, Paper I,
Fig. 2), with two stars in sample B not in sample A, and with 9 stars
in sample A not in sample B. This of course is the region which makes
the dominant contribution to the sample incompleteness and
contamination, and the different classifications may be explained by
scatter across the selection boundaries. The reason for the asymmetry
in the numbers scattered (i.e. 2 and 9) is not clear.

To reduce the uncertainty in this region, and so minimise the
contamination by A/BS stars, we combined the classifications in the
following way. For each of the 94 stars, and for each classifier, we
determined the probability the star is BHB, $p(BHB)$, by computing the
two--dimensional pdf in each diagram, given the measured values of the
parameters and the errors, and measuring the fraction below the
classification boundary. For each star, we took the average value for
the two methods, $\bar{p}(BHB)$, and classified as BHB all stars for
which $\bar{p}(BHB)>0.5$, i.e. we optimally weighted the information
in the two plots. With this procedure, 45 of the 94 stars are
classified BHB. Of the three stars in sample B not in sample A, 2 are
classified BHB. Of the 9 stars in sample A, not in sample B, 5 are
classified BHB. Finally of the 48 additional stars only possessing
spectra, without accurate photometry, 15 are classified BHB by the
{\em Scale width--Shape} method.

\begin{figure}
\rotatebox{0}{
\centering{
\scalebox{0.30}{
\includegraphics*[-50,140][700,700]{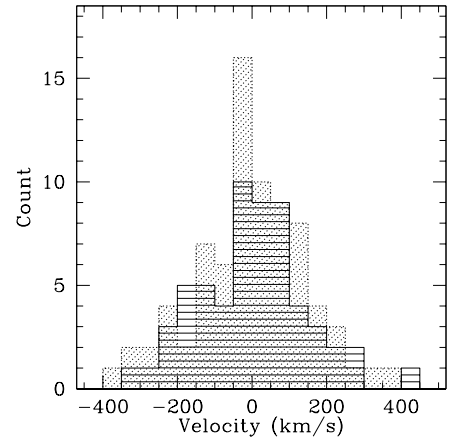}
}}}
\caption{Histograms of heliocentric radial velocities for the BHB
sample (lines) and the A/BS sample (dots). Bins are 40 km s$^{-1}$ in width.\label{histo_bs_bhb}}
\end{figure}

\begin{table*}   
\begin{center}
\begin{small}
\begin{tabular}{lrrrrrrrr}
\hline
\noalign{\smallskip}   
\multicolumn{1}{c}{Star} &   
\multicolumn{1}{c}{$l$} &     
\multicolumn{1}{c}{$b$} &   
\multicolumn{1}{c}{V$_{\odot}$} &   
\multicolumn{1}{c}{$\sigma_{\mathrm{V}}$} & 
\multicolumn{1}{c}{$d_{\odot}$} &
\multicolumn{1}{c}{$\sigma_{\mathrm{d}}$} &
\multicolumn{1}{c}{$V_{gal}$} &
\multicolumn{1}{c}{$d_{gal}$} \\
\multicolumn{1}{c}{} &   
\multicolumn{1}{c}{[$^{\circ}$]} &     
\multicolumn{1}{c}{[$^{\circ}$]} &   
\multicolumn{1}{c}{[km s$^{-1}$]} &   
\multicolumn{1}{c}{[km s$^{-1}$]} & 
\multicolumn{1}{c}{[kpc]} &
\multicolumn{1}{c}{[kpc]} &
\multicolumn{1}{c}{[km s$^{-1}$]} &
\multicolumn{1}{c}{[kpc]} \\
\noalign{\smallskip}   
\hline  
\noalign{\smallskip}  
CSGP-017   &   90.8  &   -88.4  &    27.2  &    16.1 & 50.2   &  4.1  &    26.6 &  50.8    \\ 
CSGP-021   &   54.2  &   -87.9  &   -89.9  &    15.2 & 50.6   &  4.2  &   -89.8 &  51.1    \\ 
CSGP-029   &  223.2  &   -88.5  &   -68.4  &    10.1 & 45.9   &  3.2  &   -79.8 &  46.7    \\ 
CSGP-034   &  281.9  &   -89.6  &   -41.2  &    13.6 & 51.0   &  4.0  &   -49.8 &  51.6    \\ 
CSGP-050   &  270.8  &   -86.5  &   103.9  &    13.7 & 47.0   &  3.8  &    82.7 &  47.7    \\ 
CSGP-063   &  331.2  &   -86.7  &  -102.1  &    14.0 & 42.1   &  3.2  &  -115.2 &  42.5    \\ 
CSA94-008  &  178.2  &   -50.0  &  -201.3  &    24.1 & 26.5   &  2.2  &  -207.7 &  32.2    \\ 
CSA94-010  &  176.8  &   -49.6  &    80.6  &    14.0 & 27.3   &  2.3  &    77.7 &  33.0    \\ 
CSA94-011  &  179.4  &   -51.2  &    -7.3  &    12.1 & 20.0   &  1.6  &   -16.8 &  25.8    \\ 
CSA94-013  &  177.2  &   -50.1  &  -155.3  &    12.0 & 21.3   &  1.7  &  -159.2 &  27.1    \\ 
CSA94-017  &  175.7  &   -49.6  &    11.5  &    13.0 & 17.3   &  1.3  &    11.7 &  23.3    \\ 
CSA94-036  &  173.8  &   -50.1  &    -0.9  &    14.6 & 12.2   &  0.9  &     4.0 &  18.4    \\ 
CSA94-038  &  175.4  &   -51.3  &    57.3  &    13.1 & 16.7   &  1.3  &    57.9 &  22.6    \\ 
CSA94-050  &  171.3  &   -49.9  &  -156.4  &    13.2 & 25.3   &  1.9  &  -144.8 &  31.0    \\ 
CSA94-056  &  171.5  &   -50.7  &   -13.4  &    12.5 & 11.0   &  0.8  &    -2.7 &  17.2    \\ 
CSA94-061  &  170.2  &   -50.2  &  -112.9  &    14.7 & 33.2   &  2.8  &   -98.8 &  38.7    \\ 
CSA94-072  &  172.2  &   -52.4  &  -157.3  &    13.2 & 15.7   &  1.2  &  -148.9 &  21.5    \\ 
CF358-012  &  237.9  &   -53.8  &    95.3  &    13.1 & 17.7   &  1.3  &   -29.4 &  21.6    \\ 
CF854-018  &  243.7  &    46.3  &   262.1  &    12.5 & 10.5   &  0.7  &   120.7 &  15.0    \\ 
CF854-021  &  242.6  &    46.3  &    26.3  &    11.7 & 13.7   &  1.0  &  -113.9 &  17.9    \\ 
CF854-028  &  241.2  &    45.6  &   237.9  &    15.3 & 33.0   &  3.0  &    97.6 &  36.5    \\ 
CF854-036  &  244.6  &    42.8  &   -15.4  &    17.6 & 44.0   &  3.5  &  -167.2 &  47.1    \\ 
CF854-038  &  245.5  &    41.7  &    63.8  &    15.3 & 16.4   &  1.2  &   -92.0 &  20.4    \\ 
CF854-039  &  242.0  &    43.8  &    28.8  &    15.4 & 47.8   &  4.3  &  -117.2 &  51.1    \\ 
CF854-040  &  241.4  &    44.2  &    76.0  &    11.8 & 12.9   &  1.0  &   -68.4 &  17.4    \\ 
CF854-041  &  243.9  &    42.7  &   428.6  &    14.1 & 11.4   &  0.8  &   277.1 &  15.9    \\ 
CF854-047  &  244.6  &    41.8  &    64.1  &    12.0 & 13.5   &  1.0  &   -90.4 &  17.8    \\ 
CF789-010  &  302.5  &    59.0  &   102.3  &    13.4 & 14.4   &  1.2  &    10.1 &  14.4    \\ 
CF789-022  &  301.0  &    55.3  &   280.6  &    21.3 & 28.6   &  2.2  &   175.9 &  27.3    \\ 
CF789-024  &  300.8  &    55.6  &   156.7  &    17.3 & 13.1   &  1.2  &    52.3 &  13.2    \\ 
CF789-041  &  299.4  &    59.5  &    -1.9  &    12.0 & 18.2   &  1.5  &   -96.3 &  18.0    \\ 
CF789-045  &  299.1  &    58.6  &   132.1  &    16.4 & 15.1   &  0.9  &    34.7 &  15.2    \\ 
CF789-050  &  299.0  &    57.3  &   -20.3  &    12.3 & 11.0   &  0.8  &  -121.6 &  11.8    \\ 
CF789-078  &  296.8  &    55.6  &    53.0  &    13.7 & 36.4   &  2.8  &   -56.0 &  35.2    \\ 
CF789-079  &  296.4  &    56.9  &    24.8  &    13.0 & 14.4   &  1.1  &   -80.5 &  14.7    \\ 
CF789-090  &  295.9  &    54.5  &    28.8  &    12.8 & 13.6   &  1.0  &   -84.5 &  13.9    \\ 
CF789-100  &  294.5  &    58.7  &   -56.4  &    13.9 & 12.7   &  0.9  &  -158.2 &  13.5    \\ 
CF789-108  &  294.3  &    55.4  &    38.2  &    12.1 &  9.9   &  0.7  &   -73.9 &  11.2    \\ 
CMTF-004   &   41.1  &   -52.3  &   -37.5  &    11.0 & 39.9   &  4.8  &    54.4 &  36.9    \\ 
CMTF-010   &   40.8  &   -51.2  &  -216.1  &     9.8 & 22.4   &  1.4  &  -122.4 &  19.9    \\ 
CMTF-011   &   39.2  &   -51.4  &    76.8  &    14.6 & 30.6   &  2.6  &   167.2 &  27.6    \\ 
CMTF-013   &   38.3  &   -50.3  &  -215.2  &    14.8 & 30.7   &  2.2  &  -124.0 &  27.6    \\ 
CMTF-031   &   38.4  &   -48.3  &  -130.6  &    12.8 & 43.0   &  5.6  &   -35.3 &  39.4    \\ 
CMTF-038   &   36.8  &   -53.5  &    27.4  &    13.4 & 33.3   &  2.8  &   108.7 &  30.3    \\ 
CMTF-048   &   35.8  &   -52.2  &  -190.4  &    15.8 & 25.6   &  2.0  &  -108.3 &  22.7    \\ 
CMTF-049   &   35.4  &   -52.2  &    15.2  &    14.9 & 31.7   &  2.5  &    96.6 &  28.6    \\ 
CMTF-059   &   37.3  &   -51.0  &    99.7  &     9.8 & 36.1   &  5.3  &   187.1 &  32.8    \\ 
CMTF-061   &   37.0  &   -51.0  &   -36.1  &    15.6 & 50.7   &  5.7  &    50.9 &  47.2    \\ 
CMTF-064   &   37.4  &   -50.6  &  -305.4  &    15.7 & 33.8   &  3.4  &  -216.8 &  30.6    \\ 
CMTF-073   &   34.8  &   -50.1  &   -46.5  &    13.0 & 23.3   &  1.8  &    37.8 &  20.3    \\ 
CMTF-075   &   35.7  &   -49.6  &  -285.4  &    14.1 & 30.4   &  3.5  &  -198.2 &  27.1    \\ 
CMTF-079   &   35.2  &   -49.2  &  -178.6  &    13.2 & 26.9   &  2.1  &   -91.8 &  23.6    \\ 
CMTF-102   &   34.0  &   -51.5  &   132.1  &    14.8 & 26.9   &  3.7  &   212.1 &  23.8    \\ 
CMTF-123   &   33.0  &   -49.1  &  -104.6  &     8.8 & 31.2   &  3.1  &   -22.2 &  27.6    \\ 
CMTF-128   &   31.4  &   -49.1  &  -124.9  &    14.3 & 23.8   &  3.1  &   -46.0 &  20.4    \\ 
\noalign{\smallskip}   
\hline 
\end{tabular}
\caption{Summary of kinematic information for the candidates
classified as blue horizontal branch stars.\label{tab_data_sum}}    
\end{small}
\end{center}
\end{table*}

\begin{table*}   
\begin{center}
\begin{small}
\begin{tabular}{lrrrrrrrr}
\hline
\noalign{\smallskip}   
\multicolumn{1}{c}{Star} &   
\multicolumn{1}{c}{$l$} &     
\multicolumn{1}{c}{$b$} &   
\multicolumn{1}{c}{V$_{\odot}$} &   
\multicolumn{1}{c}{$\sigma_{\mathrm{V}}$} & 
\multicolumn{1}{c}{$d_{\odot}$} &
\multicolumn{1}{c}{$\sigma_{\mathrm{d}}$} &
\multicolumn{1}{c}{$V_{gal}$} &
\multicolumn{1}{c}{$d_{gal}$} \\
\multicolumn{1}{c}{} &   
\multicolumn{1}{c}{[$^{\circ}$]} &     
\multicolumn{1}{c}{[$^{\circ}$]} &   
\multicolumn{1}{c}{[km s$^{-1}$]} &   
\multicolumn{1}{c}{[km s$^{-1}$]} & 
\multicolumn{1}{c}{[kpc]} &
\multicolumn{1}{c}{[kpc]} &
\multicolumn{1}{c}{[km s$^{-1}$]} &
\multicolumn{1}{c}{[kpc]} \\
\noalign{\smallskip}   
\hline  
\noalign{\smallskip}  
IF358-25   &  237.8  &   -54.9  &   162.4  &    13.0 & 43.2   &  3.5  &    40.9  & 46.3   \\
IF854-008  &  239.6  &    44.4  &   165.8  &    12.1 & 12.1   &  1.0  &    24.6  & 16.7   \\
IF854-058  &  243.9  &    41.8  &   173.4  &    16.8 & 38.7   &  3.7  &    19.9  & 42.0   \\
IMTF-17    &   31.6  &   -50.6  &  -164.4  &    16.6 & 32.1   &  3.2  &   -87.8   &28.6    \\
IMTF-37    &   35.0  &   -50.6  &   -13.3  &    13.5 & 42.2   &  2.9  &    70.5   &38.6    \\
\noalign{\smallskip}   
\hline 
\end{tabular}
\caption{Data summary for the horizontal branch stars contd.}   
\end{small}
\end{center}
\end{table*}

In total there are 60 stars classified as BHB, and Table
\ref{tab_data_sum} gives a summary of their kinematic properties. The
contamination of this sample by A/BS stars will be approximately
$10\%$.  Note that the degree of contamination is not a function of
distance, since we increased the exposure times to maintain an
approximately constant level of S/N in our sample. Listed in Table
\ref{tab_data_sum} are the Galactic coordinates $l$ and $b$, the
heliocentric velocity, V$_{\odot}$, its 1$\sigma$ error
$\sigma_{\mathrm{V}}$, and the heliocentric distance $d_{\odot}$, and
its 1$\sigma$ error $\sigma_{\mathrm{d}}$.  The last two columns
provide the Galactocentric radial velocity and distance, V$_{gal}$ and
d$_{gal}$ respectively.  To convert the heliocentric quantities to
Galactocentric quantities the heliocentric radial velocities are first
corrected for solar motion by assuming a solar peculiar velocity of
($U,V,W$) = (-9,12,7), where $U$ is directed outward from the Galactic
Centre, $V$ is positive in the direction of Galactic rotation at the
position of the sun, and $W$ is positive towards the North Galactic
Pole. We have assumed a circular speed of 220 km s$^{-1}$, at the
Galactocentric radius of the sun ($R_{\odot}$ = 8.0 kpc). This table
is the main result of the paper, and will be used in our forthcoming
dynamical study.

Figure \ref{histo_bs_bhb} shows histograms of the line-of-sight
heliocentric radial velocity of the BHB and A/BS samples.  Little
difference in the spread of heliocentric radial velocities for the two
samples is evident.  A Kolmogorov-Smirnov test gives a probability of
0.97 that the blue stragglers and BHB distributions are drawn from the
same population.

It is of interest to compare the measured velocity dispersion of our
sample of 60 BHB stars, mean distance 28 kpc, with the result of
Norris \& Hawkins (1991, hereafter NH) for a small sample of 9 remote
halo BHB stars, mean distance 55 kpc. After quadratically subtracting
the measurement errors, in the same manner as NH, the measured
dispersion of the radial component of the Galactocentric velocity for
our BHB sample is 108 $\pm$ 10 km s$^{-1}$ (the result for the blue
stragglers is 115 $\pm$ 10 km s$^{-1}$). NH measured a value of 111
$\pm$ 25 km s$^{-1}$, in excellent agreement. Our value is also
consistent with results by Sommer-Larsen et al. 1997 who find that the
velocity dispersion falls from $\sigma_{r}$ = 140 $\pm$ 10 km s$^{-1}$
in the solar neighbourhood, to an asymptotic value of 89 $\pm$ 19 km
s$^{-1}$ at R $\geq$ 20 kpc. More recently, Sirko et al. 2004a isolate
large samples of distant BHB stars using the Sloan Digital Sky
Survey. They split their sample into a bright ($g<18$) subsample,
which is contaminated by blue stragglers at the level of about 10\%
(i.e. similar to the work presented here) and a faint subsample
($g>18$), which is contaminated at about 25\%. If we consider only
their bright sample then $\sigma_{r}$ = 99.4 $\pm$ 4.3 (Sirko et
al. 2004b), at a mean distance 16 kpc, again consistent with our
value. It is worth noting for completeness that the inclusion of their
faint sample changes their result very little at $\sigma_{r}$ = 101.4
$\pm$ 2.8.

\section{Summary}

In this paper we have presented the results of a survey of remote halo
A-type stars, beginning with catalogues of $ub_jr$ photometry from
UKST plates in two northern and four southern high-Galactic-latitude
fields.  Accurate CCD photometry and spectroscopy of candidate A-type
stars produced a sample of 142 stars with data of suitable quality for
classification into the classes BHB and A/BS. The final sample, Table
14, comprises 60 stars classified BHB, at distances of $11-52$kpc from
the Sun (mean distance 28 kpc), with heliocentric radial velocities
accurate to 15 km s$^{-1}$, and distance errors $<10\%$. These stars
are suitable for a dynamical study of the mass distribution in the
Milky Way at large radii. The stars are all metal poor, with mean
[Fe/H]=-1.8 with dispersion 0.5. The measured dispersion of the radial
component of the Galactocentric velocity for this sample is 108 $\pm$
10 km s$^{-1}$ which is consistent with Sommer-Larsen et al. 1997 and
Sirko et al. 2004b. The remaining 82 stars are classified A/BS. These
stars have very similar metallicities and velocity dispersion to the
BHB stars.

\end{document}